\documentclass[twocolumn]{aastex701}

\newcommand\xmm{\textit{XMM-Newton}}
\newcommand\chandra{\textit{Chandra}}

\newcommand\nustar{\textit{NuSTAR}}
\newcommand\suzaku{\textit{Suzaku}}

\newcommand\xrism{\textit{XRISM}}
\newcommand\nicer{\textit{NICER}}

\newcommand\ixpe{\textit{IXPE}}

\newcommand\pcm{cm$^{-2}$}

\usepackage{multirow}

\begin{document}

\title{\xrism/Resolve observations of Hercules X-1: a pulsating, highly broadened Fe K emission line from the neutron star accretion column}

\author[orcid=0000-0003-4511-8427,sname='Kosec']{Peter Kosec}
\affiliation{Center for Astrophysics | Harvard \& Smithsonian, Cambridge, MA, USA}
\email[show]{peter.kosec@cfa.harvard.edu}

\author[orcid=0000-0003-2663-1954,sname='']{Laura Brenneman}
\affiliation{Center for Astrophysics | Harvard \& Smithsonian, Cambridge, MA, USA}
\email[]{lbrenneman@cfa.harvard.edu}

\author[orcid=0000-0003-0172-0854,sname='']{Erin Kara}
\affiliation{MIT Kavli Institute for Astrophysics and Space Research, Massachusetts Institute of Technology, Cambridge, MA 02139, USA}
\email[]{ekara@space.mit.edu}

\author[orcid=0000-0003-2532-7379,sname='']{Ciro Pinto}
\affiliation{INAF—IASF Palermo, Via U. La Malfa 153, I-90146 Palermo, Italy}
\email[]{ciro.pinto@inaf.it}

\author[orcid=0000-0002-5359-9497,sname='']{Daniele Rogantini}
\affiliation{Department of Astronomy and Astrophysics, The University of Chicago, Chicago, IL 60637, USA}
\email[]{danieler@uchicago.edu}

\author[orcid=0000-0003-1498-1543,sname='']{Rüdiger Staubert}
\affiliation{Institut für Astronomie und Astrophysik, Universität Tübingen, Sand 1, D-72076 Tübingen, Germany}
\email[]{staubert@astro.uni-tuebingen.de}

\author[orcid=0000-0001-5819-3552,sname='']{Dominic Walton}
\affiliation{Centre for Astrophysics Research, University of Hertfordshire, College Lane, Hatfield AL10 9AB, UK}
\email[]{dwalton354@gmail.com}

\author[orcid=0000-0001-5852-6740,sname='']{Francesco Barra}
\affiliation{Università degli Studi di Palermo, Dipartimento di Fisica e Chimica, via Archirafi 36, I-90123 Palermo, Italy}
\email[]{francesco.barra@inaf.it}

\author[orcid=0000-0002-9378-4072,sname='']{Andrew Fabian}
\affiliation{Institute of Astronomy, Madingley Road, Cambridge, CB3 0HA, UK}
\email[]{acf@ast.cam.ac.uk}

\author[orcid=0000-0003-1244-3100,sname='']{Teruaki Enoto}
\affiliation{Department of Physics, Graduate School of Science, Kyoto University, Kitashirakawa Oiwake-cho, Sakyo-ku, Kyoto, 606-8502, Japan}
\email[]{enoto.teruaki.2w@kyoto-u.ac.jp}

\author[orcid=0000-0003-2869-7682,sname='']{Jon M. Miller}
\affiliation{Department of Astronomy, University of Michigan, Ann Arbor, MI 48109, USA}
\email[]{jonmm@umich.edu}

\author[orcid=0009-0006-7889-6144,sname='']{Takuto Narita}
\affiliation{Department of Physics, Graduate School of Science, Kyoto University, Kitashirakawa Oiwake-cho, Sakyo-ku, Kyoto, 606-8502, Japan}
\email[]{narita.takuto.43e@st.kyoto-u.ac.jp}

\author[orcid=0009-0007-8032-3641,sname='']{Koh Sakamoto}
\affiliation{Department of Physics, Graduate School of Science, Kyoto University, Kitashirakawa Oiwake-cho, Sakyo-ku, Kyoto, 606-8502, Japan}
\email[]{sakamoto.kou.68f@st.kyoto-u.ac.jp}

\author[orcid=0009-0003-9261-2740,sname='']{Yutaro Nagai}
\affiliation{Department of Physics, Graduate School of Science, Kyoto University, Kitashirakawa Oiwake-cho, Sakyo-ku, Kyoto, 606-8502, Japan}
\email[]{nagai.yuutarou.25r@st.kyoto-u.ac.jp}




\begin{abstract}

The study of X-ray pulsar accretion columns helps us characterize accretion physics in this extreme regime of strong gravity and strong magnetic fields. Previous observations of the X-ray pulsar Hercules X-1 revealed a highly broadened Fe K emission line, associated with Doppler motions exceeding 0.1c, suggesting its origin in the accretion column. We obtained a high-spectral resolution view of the Fe K energy band of Hercules X-1 thanks to a 200 ks observation with the \xrism\ observatory. The \xrism/Resolve microcalorimeter spectra allow us to separate the different spectral components and accurately model them with phenomenological models. We confirm the presence of a broad line near 6.5 keV with a typical $1\sigma$ width of 1 keV. Performing a pulse-phase-resolved analysis, we find that the feature is strongly variable with Her X-1 pulse phase. This is consistent with the proposed origin due to collisional recombination or by reprocessing of the primary X-ray emission in the accretion column, where strong variability with pulse phase is expected due to the rotation of the columns alongside with the neutron star. Additionally, the Fe K line pulsation pattern evolves with the 35-day cycle of Hercules X-1, supporting the scenario that the neutron star and its accretion columns undergo precession, in agreement with recent polarimetric results from the \ixpe\ observatory. We discuss the future applications of modeling of this broad line in X-ray pulsars with physical spectral models. This could be used to detect and track neutron star precession, advancing our understanding of neutron star interiors.

\end{abstract}



\section{Introduction}

In an X-ray binary powered by accretion onto a highly magnetized neutron star (B$~\gtrsim10^{10}$ G), the neutron star's magnetic field disrupts the inner part of the accretion disk, channeling the infalling matter along the magnetic field lines and onto the magnetic poles \citep{Lamb+73}. The radius of the magnetosphere (the region within which the disk is disrupted) depends on the strength of the magnetic field as well as the mass accretion rate through the disk \citep{Ghosh+79}. The strong deceleration of the accreting matter close to the neutron star surface produces intense X-ray emission \citep{Becker+07}, and at high mass accretion rates a structure called the accretion column is formed \citep{Basko+76}. As the whole magnetosphere including the column co-rotates along with neutron star and the column emission pattern is anisotropic, the X-ray emission of these systems is periodically pulsed and they are known as X-ray pulsars \citep{Giacconi+71}. Thanks to their relatively common occurrence in our Galaxy and isotropic luminosities reaching as high as $10^{37-38}$ erg/s, X-ray pulsars and the X-ray emission of their accretion columns have been studied in detail using both phenomenological as well as physical spectral models \citep[for a recent review, see][]{Mushtukov+22}. The X-ray spectra of these objects are dominated by bulk and thermal Comptonization in the accretion column, but their smooth nature gives us limited information about the physical properties, location, size and the orientation of the column.

Recently, we proposed a new way to study the accretion column properties \citep{Kosec+22}. We studied the X-ray spectrum of the classical X-ray pulsar Hercules X-1 \citep[hereafter Her X-1,][]{Tananbaum+72} using \xmm\ and \chandra\ and detected a highly broadened Fe K emission line with a best-fitting full width at half maximum (FWHM) ranging from 1.5 to 3 keV ($1\sigma$ width of 0.6 to 1.3 keV). If the width of the feature originates from Doppler velocity broadening, very high velocities of $0.1-0.2$c are required. This is important in Her X-1 because its magnetic field is accurately measured via its cyclotron resonance scattering feature energy and is $\sim4\times10^{12}$ G \citep{Staubert+19}. Therefore, the accretion disk should be disrupted by the magnetosphere far away from the neutron star and such high velocities should not be present in the disk of Her X-1. A Keplerian velocity of more than 0.1c is only present in the inner 100 $R_G$ of an accretion disk ($R<50$ R$_{G}$ is required for velocities approaching $v\sim0.2$c), but the field of Her X-1, assuming a dipole, should disrupt the disk at $\sim1000~R_G$ \citep{Ghosh+78}. If the field of Her X-1 is dominated by a more complex, multi-polar component, the magnetosphere size may be smaller, increasing the range of possible inner disk Keplerian velocities, but a decrease of magnetosphere size by a factor of more than 10 is unlikely \citep{Brice+21}. We note that the model of \citet{Scott+00} based on the Her X-1 pulse shape evolution over the course of the 35-day warped disk precession cycle \citep{Gerend+76} predicts a much smaller inner disk radius of $\sim200$ R$_{G}$, but even that is far larger than required to produce the Fe K line of the observed width using Keplerian motion. Therefore, the Fe K emission line is too broad to originate in the accretion disk, and its most likely origin is in the accretion column of the neutron star.

Fe K lines are commonly detected in X-ray pulsars and other X-ray binaries, and this broad feature was observed in \nustar\ and \suzaku\ observations of Her X-1 by \citet{Fuerst+13} and \citet{Asami+14}. However, in \citet{Kosec+22} we interpreted this feature as an emission line originating directly in the accretion column for the first time. To first order, the line could be reproduced with a single Gaussian in the moderate resolution datasets, indicating that it is not strongly asymmetric, e.g. it does not have a strongly red-skewed shape as seen in relativistic X-ray reflection \citep[e.g.][]{Fabian+89, Tanaka+95}. 

At the same time, in \citet{Kosec+22, Kosec+23a} we also showed how complex the Fe K energy band of Her X-1 is. The Fe K band emission is composed of the highly broadened Fe K line discussed above, as well as a medium-width ($\sim0.5$ keV FWHM, plausibly originating near the magnetosphere boundary) Fe XXV line and much narrower Fe I K$\alpha$ and K$\beta$ lines (originating in the outer accretion disk). The accretion disk wind of Her X-1 \citep{Kosec+20, Kosec+24} additionally imprints highly ionized, variable Fe XXV and XXVI absorption features on top of these emission lines. Therefore, an observation with a high-spectral resolution X-ray instrument covering the entire Fe K energy range and its vicinity ($\sim4-9$ keV) is required to fully resolve the individual spectral components and accurately measure the properties of the highly broadened line.

\begin{figure*}
\begin{center}
\includegraphics[width=0.85\textwidth]{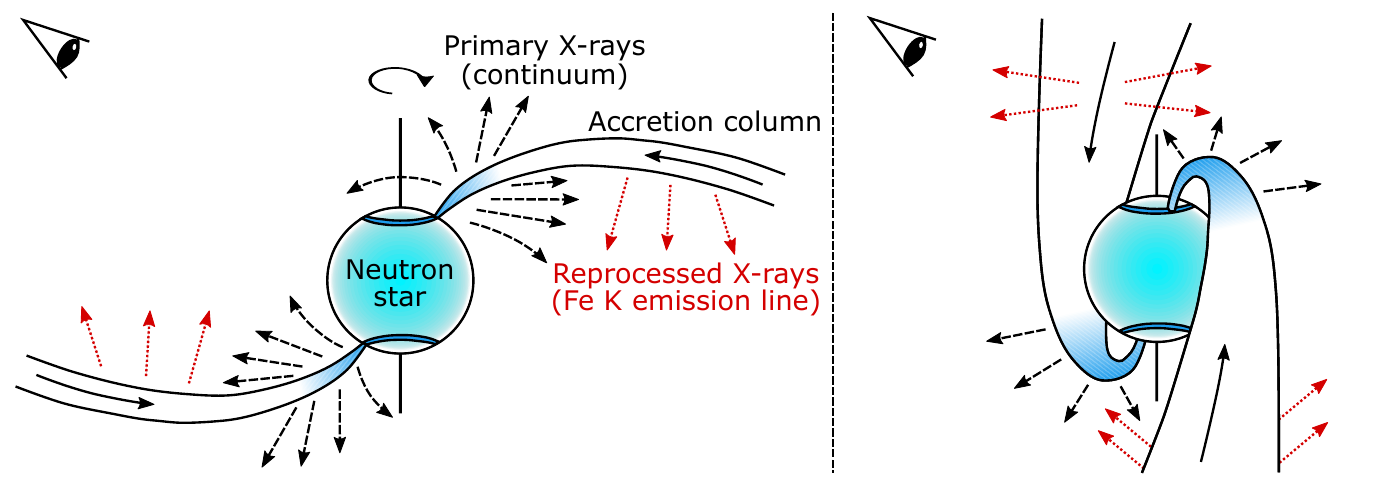}
\caption{A schematic of Her X-1, its accretion column and the possible emission sites of the broad Fe K line. Accreting matter follows magnetic field lines along the column (solid arrows). As it decelerates, it emits primary X-rays (dashed arrows). Some of the primary emission is reprocessed by the column and the Fe K line emission is produced (maroon dotted arrows). As the star rotates, due to the changes in the observed accretion column solid angle, projected infall velocity, relativistic beaming and light bending, the Fe line flux, energy and width should all vary with the neutron star rotation phase. \label{Pe_HerX1_scheme}}
\end{center}
\end{figure*}

Such a highly broadened Fe K line may be produced by recombination in the accretion column or by reprocessing of the primary column emission off the different parts of the column itself. A simple schematic of this latter situation is shown in Fig. \ref{Pe_HerX1_scheme}. A crucial prediction of this hypothesis is that since both the magnetosphere and the accretion column periodically rotate, if the line originates in the column, it should also be pulsed at the rotation frequency of the neutron star. The variation in the line properties with pulse phase will happen due to a combination of the changes in the observed column solid angle and the Fe K reprocessing area from our line of sight, changes in the projected column infall velocity and due to relativistic beaming and light bending. Importantly, not only the line flux should pulsate, but also its energy and width.

Fe K emission lines have been previously shown to pulsate in a number of X-ray pulsars including V 0332+53, Swift J0243.6+6124 and Cen X-3 \citep{Bykov+21, Xiao+24, Roy+25}, but the line was not explicitly discussed to originate in the accretion column of these systems. Fe K line emission in Her X-1 was previously studied with pulse-phase-resolved analyses by \citet{Ramsay+02} and by \citet{Vasco+13} and indeed appears to be pulsed, but the Fe K energy band emission was treated as a single (Gaussian) line component in those studies, without treating the individual spectral components described above separately.

The Fe K band of Her X-1 was recently resolved in detail using a 200 ks high-spectral resolution observation by the recently launched \xrism\ observatory \citep{Kosec+26}. Here we perform a time-averaged and pulse-phase-resolved analysis of the broad Fe K line in Her X-1, combining this high-quality dataset with the full spectral model of the Her X-1 Fe K energy band from that work. The resolution of \xrism\ allows us to accurately separate all the individual spectral components and track their evolution, providing the highest-to-date fidelity determination of the broad Fe K line variation with pulse phase.

The structure of this manuscript is as follows. In section \ref{sec:obsdataprep}, we describe our data reduction and preparation methods. In section \ref{sec:results}, we describe the spectral model adopted, and show the results of our analysis. Section \ref{sec:discussion} presents further discussion of the results and we conclude in section \ref{sec:conclusions}. Appendices \ref{app:orb1_orb3} and \ref{app:narrowband} contain further checks of our analysis and additional results. We assume a Her X-1 distance of 6.1 kpc \citep{Leahy+14}.

\section{Observations and Data Preparation}
\label{sec:obsdataprep}

Her X-1 was observed using \xrism\ \citep{Tashiro+22} starting from September 10 2024 for a gross duration of 380 ks, resulting in a clean exposure of 210 ks. In addition to \xrism\ data, we also obtained coordinated observations with \xmm\ \citep[][80 ks exposure]{Jansen+01}, \chandra\ \citep[][50 ks exposure]{Weisskopf+00} and \textit{Nuclear Spectroscopic Telescope Array} \citep[\nustar,][40 ks exposure]{Harrison+13}. This large observational campaign and the individual pointings are described in detail in section 2.1 of \citet{Kosec+26}. The \xrism\ observation spanned nearly 3 full Her X-1 orbital periods (P=1.7 days) during which Her X-1 was in the Main High state \citep{Katz+73}. For a lightcurve of the full \xrism\ observation, see Fig. 1 of \citet{Kosec+26}. Following \citet{Kosec+26}, we name these 3 orbits Orbit 1, 2 and 3. During Orbits 2 and 3, there are \nustar\ snapshots ($\sim20$ ks exposure each), which are simultaneous with part of the \xrism\ coverage. For Orbit 1, there is no simultaneous \nustar\ coverage.

To perform an in-depth analysis of the highly broadened Fe K emission line, we use all available \xrism\ Main High state data. As the feature originates close to the neutron star, it is not detected in the Low, dipping or eclipse states of Her X-1 when the inner accretion flow is obscured. We therefore exclude all periods of low state, absorption dips and eclipses. We use all 3 Her X-1 orbits but do not stack them into a single dataset, as significant evolution is observed in the spectral shape and the flux of Her X-1, as well as in the best-fitting Her X-1 rotation period. Additionally, the excellent signal-to-noise (S/N) offered by \xrism\ data allows us to reach sufficient photon statistics using individual orbit datasets. Hence, we use this opportunity to investigate how the pulse-phase-resolved properties of the Fe K line vary over time, and particularly how they change with the increasing precession phase of the 35-day cycle of Her X-1. The dataset details for all 3 orbits are given in Table \ref{orbit_data_table}.

\begin{table*}
\caption{Details of the \xrism\ and \nustar\ observations used in this work and their splitting into different Her X-1 orbits. For \nustar\ observations, we give the clean exposure of the FPMA instrument. For \xrism\ observations, the clean exposure of the Resolve instrument is given. \label{orbit_data_table}}
\begin{tabular}{ccccccc}
\hline
\hline
Orbit & Observatory & Observation ID & Start Time & End Time & Exposure Time & Best-fitting period\\
&  &&ISO 8601 date  &ISO 8601 date & s  & s \\
&&&MJD&MJD&&\\
\hline
1 & \xrism & 201074010 &2024-09-10 11:55 & 2024-09-11 11:51& 40077 & 1.2376982(2)\\
&&&60563.497&60564.494&&\\
2 & \xrism & 201074010 & 2024-09-11 21:24  & 2024-09-13 03:43 & 60515  & 1.2376973(2) \\
&&&60564.892&60566.155&&\\
 & \nustar & 81002350002 & 2024-09-11 21:51 & 2024-09-12 08:51 & 17953    &  \\
&&&60564.910&60565.369&&\\
3  & \xrism  & 201074010 &  2024-09-13 14:03 & 2024-09-14 10:45 &  43859 & 1.2376969(2) \\ 
&&&60566.585&60567.448&&\\
 & \nustar &81002350004 & 2024-09-13 13:51 & 2024-09-14 00:56   & 18120 & \\
&&&60566.577&60567.039&&\\
\hline
\end{tabular}
\end{table*}

In this analysis, we use \xrism/Resolve and \nustar\ (FPMA and FPMB) data to constrain the broadband $1.8-75$ keV spectrum and to characterize the broad line properties. The data reduction procedures for both datasets are described in detail in sections 2.2 and 2.4 of \citet{Kosec+26}. We did not use any \xrism/Xtend data in this study due to the inferior spectral resolution compared with the Resolve instrument in the Fe K energy band, as well as due to possible pile-up in these data. 

After obtaining individual cleaned event files for each Her X-1 orbit, we performed a pulsation search using the \xrism\ event file to determine the best-fitting Her X-1 frequency for that orbit. Afterwards, we extracted pulse-phase-resolved spectra used for spectral fitting.

First, in order to facilitate these pulsation searches, we corrected all photon arrival times into the Solar System barycenter using the \textsc{barycen} routine for \xrism\ event files, and using the \textsc{barycorr} routine for \nustar\ event files. Secondly, we corrected all photon arrival times in both \xrism\ and \nustar\ event files for the orbital motion of Her X-1. We used the orbital ephemeris from \citet{Staubert+09} with an orbital period of 1.700167590 days, an eclipse midpoint on MJD 46359.871940, a change of the orbital period of $-4.85 \times 10^{-11}$ s s$^{-1}$ and $a~\textrm{sin}~i$ of 13.1831 s. Afterwards, we performed a simple pulse period search around the value observed for Her X-1 by the Fermi/Gamma-ray Burst Monitor Pulsar Project \citep{Finger+09} for these dates, which was around $P=1.237697$ s. We used the \textsc{epoch\_folding\_search} function within the Stingray package \citep{Huppenkothen+19, Stingray+19} to locate the peak of the $\chi^2$ epoch folding (EF) statistics and determine the pulsation period for each Her X-1 orbit. The best-fitting values are listed in Table \ref{orbit_data_table}.

Using these timing solutions, we produced Good Time Interval (GTI) files containing the split times for pulse-phase-resolved extraction. We aim to obtain as fine time resolution in this analysis as possible, without compromising the S/N per pulse bin. To assess the dataset S/N, we performed two different pulse-phase-resolved extractions for each of the 3 Her X-1 orbits by splitting the data into 21 pulse bins and into 30 pulse bins, and performing the full pulse-phase-resolved spectral analysis (described in section \ref{sec:results}) on both extractions. Additionally, to increase the S/N in each pulse bin, the width of each bin was chosen to be 3 times the pulse bin separation. Therefore, the bin width was 1/7 of the pulse period for the coarser 21 bin pulse-phase-resolved extraction ($\sim0.18$ s), and 1/10 of the pulse period for the finer 30 bin extraction ($\sim0.12$ s). 

This strategy results in neighboring bins having partially overlapping photon events, and so they are not completely independent. This washes out sharp bin-to-bin variability as the resulting pulse profile is a convolution of the real pulse profile with the used (over-sized) bin width. However, this strategy gives us extra information as it minimizes the influence of the choice of the exact pulse bin split timing (specifically their start/stop times) with respect to the real source pulse variability. Therefore, it gives us as much time resolution (and as dense time sampling as possible) while preserving sufficient S/N ratio in each pulse bin, and the cost of some smoothing of the bin-to-bin variability. In practice, we found that this approach revealed finer pulse profile variability that would be harder to visually identify in a standard (coarser) pulse-phase-resolved analysis where the bin width is equal to 1/N of the pulse period, N being the total number of pulse bins.

Based on the size of the parameter uncertainties in spectral fitting, we found that Orbit 2 (which has the highest number of photon counts) has sufficient S/N for a 30 pulse bin analysis, but this is not the case for Orbit 1 and 3. Therefore, for consistency, in the main body of this paper we only show the analysis of the 21 bin pulse-phase-resolved analysis for all 3 Her X-1 orbits. Appendix \ref{app:narrowband} focuses on comparison of different spectral models for only the highest S/N Orbit 2, and therefore we present the finer 30 bin pulse-phase-resolved analysis in that section. Additionally, to check our spectral fitting results, we also performed a coarser pulse-phase-resolved spectral analysis. In this check we split the pulse period into 15 pulse bins (each with a width of 0.2 of pulse period) and into 12 pulse bins (each with a width of 0.25 of pulse period). We confirmed that the results of the two coarser analyses are consistent with the 21-bin analysis, except some of the variability seen in the 21-bin analysis is more washed out in the coarser analysis due to the longer exposure per pulse bin.

\begin{figure}
\begin{center}
\includegraphics[width=\columnwidth]{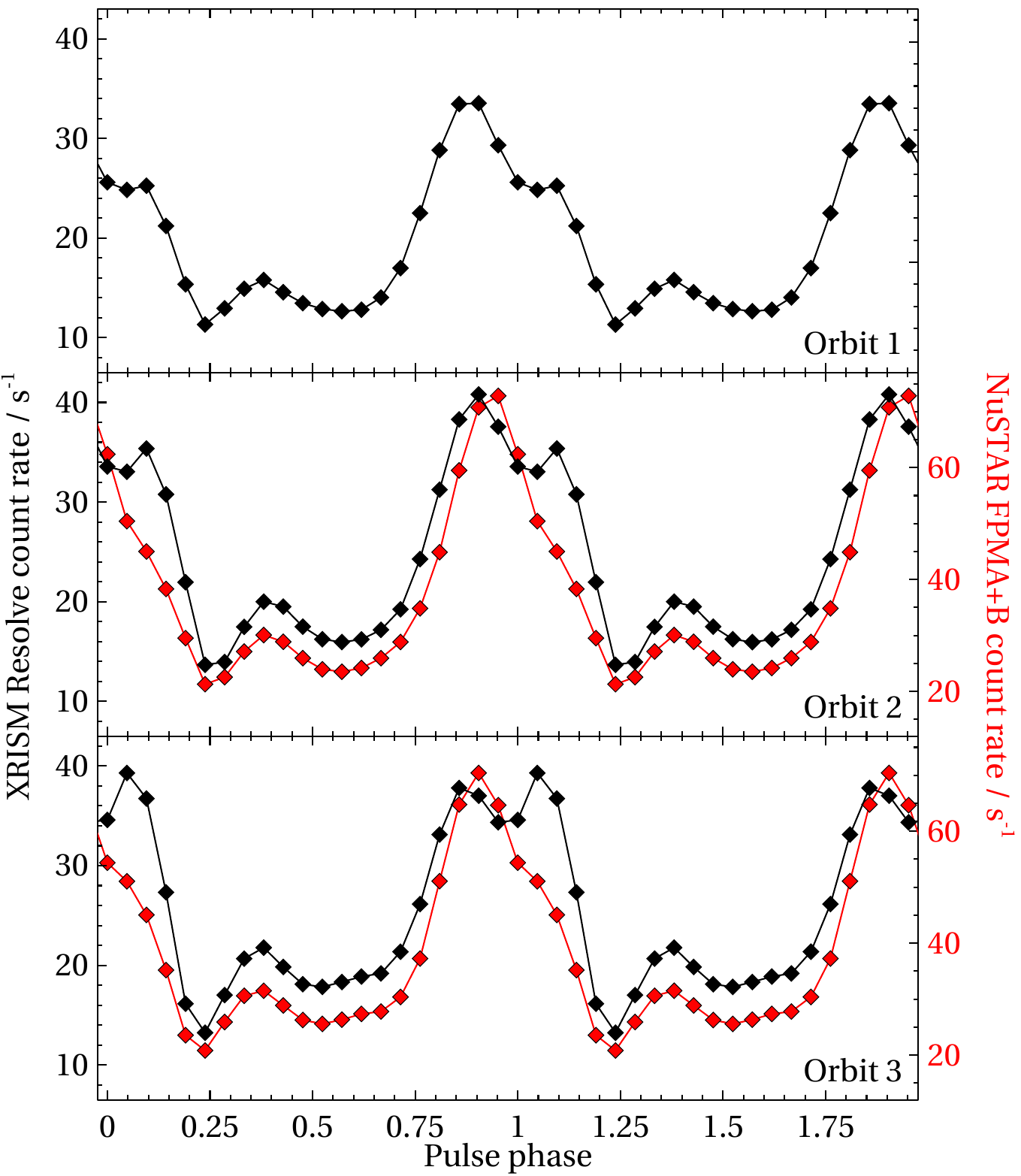}
\caption{\xrism\ (black) and \nustar\ (red) pulse profiles for all 3 Her X-1 orbits (using the 21 pulse bin extraction) in the energy bands which we used for pulse-phase-resolved spectral fitting: $1.8-12$ keV for \xrism/Resolve and $11-75$ keV for \nustar\ FPMA and FPMB. The uncertainties are smaller than the size of the datapoints. \label{Pulse_profiles}}
\end{center}
\end{figure}

The resulting pulse profiles of \xrism\ and \nustar\ data for each Her X-1 orbit, using the 21 pulse bin extraction are shown in Fig. \ref{Pulse_profiles}. Here the pulse profiles are extracted in the energy bands which were later used for pulse-phase-resolved spectral analysis, which are $1.8-12.0$ for \xrism/Resolve, and $11-75$ keV for \nustar\ FPMA and FPMB (the choice of energy bands is described in the following paragraph). We note that all \xrism/Resolve events (Hp through Ls quality) were used for this comparison, as they are all real X-ray events (background should be negligible given the high Her X-1 flux), only not all have sufficient spectral resolution for our spectral analysis. The pulse profiles from \xrism\ and \nustar\ are qualitatively different given the differing energy bands and the complex pulsation pattern of the Her X-1 continuum \citep{Fuerst+13}. As a double check, we also extracted the pulse profiles in a common energy band ($3-12$ keV) and confirmed that they are nearly identical for the same Her X-1 orbit. Only very small differences remained ($\lesssim10$\%), which are due to residual calibration differences between \xrism\ and \nustar, as well as due to the fact that the observations are not strictly simultaneous (\nustar\ observations only cover parts of the Her X-1 exposures). 

By default, the absolute value of the pulse phase does not have any useful meaning as it is given by the start time of the first GTI interval of the observing interval. We thus manually aligned the pulse profiles from the 3 Her X-1 orbits for visual purposes so that the main peak of the pulsation is roughly centered at phase equal to 0 for all 3 Her X-1 orbits.

Afterwards, we extracted \xrism/Resolve and \nustar\ FPMA and FPMB spectra for each phase bin using the generated GTI files. We note that we used the same GTI intervals for both \xrism\ and \nustar\ data to ensure a correct pulse phase extraction. For \xrism\ data, we extracted both the High-quality primary (Hp) and Medium-quality primary (Mp) events and used both, fitted simultaneously in individual spectral files in our spectral analysis. We used XL-sized response matrices for all \xrism\ datasets. Both \xrism\ and \nustar\ data were optimally binned \citep{Kaastra+16} using the \textsc{ftgrouppha} routine. The \xrism\ data were used between 1.8 and 12 keV as the gate valve was closed during the Her X-1 observation with no usable soft X-ray effective area. The \nustar\ data were used between 11 keV and 75 keV. The lower energy limit was set to retain a small overlap between the \xrism\ and \nustar\ spectra and constrain the cross-calibration constant between the two instruments. We did not use \nustar\ data between 3 and 11 keV as they have a much lower spectral resolution than Resolve data. Therefore, the individual spectral components of the Fe K band of Her X-1 cannot be accurately resolved in \nustar\ spectra alone. However, the very high photon statistics (thanks to the large effective area of \nustar) result in small uncertainties in \nustar\ spectra, which could then drive the spectral fit and systematically bias the best-fitting spectral parameters due to the insufficient spectral resolution. For this reason, we use just \xrism/Resolve data below 11 keV, to obtain an unbiased, high-resolution view of this band.

We note that for the spectral fitting of Her X-1 Orbit 1, which had no simultaneous \nustar\ exposure, we used the \nustar\ observation taken during Her X-1 orbit 2. Although they are quantitatively different, the pulse profiles of the \xrism/Resolve data (in the $1.8-12$ keV band) do not dramatically vary between the different Her X-1 orbits (Fig. \ref{Pulse_profiles}). Therefore, this approximation should not introduce significant errors in the derived Fe K energy band spectral component parameters. We do not draw any conclusions about the underlying broadband continuum, which may be more affected by this choice. Additionally, any differences in the overall Her X-1 flux between Orbit 1 and 2 will be absorbed in the cross-calibration constant used in the spectral fit between the \xrism\ and \nustar\ spectra (discussed in more detail in the following section).

\section{Spectral analysis and results}
\label{sec:results}

\subsection{Details of spectral modeling}
\label{sec:modelsetup}

All spectral fitting was performed in the \textsc{xspec} fitting package \citep{Arnaud+96} version 12.14.1. \textsc{xspec} was chosen for this analysis as it was more computationally efficient than \textsc{spex} \citep{Kaastra+96} for this analysis of many individual datasets (pulse bins). Additionally, \textsc{xspec} offers physical models of the X-ray pulsar continuum, important for constraining the shape and properties of the highly broadened Fe K emission line. Previously, we used \textsc{spex} for spectral fitting of this \xrism\ dataset in \citet{Kosec+26} as we focused on the properties of narrow absorption lines from the highly ionized disk wind. These narrow spectral features are not the focus of the present work.

All spectra are fitted by minimizing the Cash statistic \citep[C-stat,][]{Cash+79}, and uncertainties are provided at $1\sigma$ significance.  \xrism/Resolve Hp, Mp and \nustar\ FPMA and FPMB spectra were all fitted simultaneously without stacking, using cross-calibration constants. The value of the constant was fixed to 1 for the \xrism\ Hp event spectrum, and left free to vary for the remaining spectra. These constants account for any residual differences between the instrument calibrations, but also for any real variations in the Her X-1 X-ray flux over time considering that the \xrism\ and \nustar\ datasets are not strictly simultaneous (\nustar\ exposure only covers part of the \xrism\ coverage). The value for Mp is fully consistent with unity, while the cross-calibration constants for FPMA and FPMB data are typically within 10\% from unity.

To determine the broad Fe K emission line properties as accurately as possible, we use a complex X-ray spectral model to describe the 1.8-75 keV spectrum of Her X-1. The model is a combination of physical and phenomenological spectral components and is motivated by the model that we used in \citet{Kosec+26}.

The primary X-ray emission from the accretion column of Her X-1 is described using the \textsc{bwcycl} spectral model, which is based on the model by Becker \& Wolff \citep[hereafter the BW model,][]{Becker+07, Ferrigno+09, Thalhammer+21}. This model describes the X-ray spectrum of the accreting, shocked gas in the neutron star accretion column, produced primarily by bulk and thermal Comptonization, across the entire 1.8-75 keV energy band. It produces a relatively flat continuum shape with a high-energy quasi-exponential cutoff. The BW model was previously applied to a \nustar\ observation of Her X-1 by \citet{Wolff+16}. As opposed to that work, which focused on the physical interpretation of the best-fitting accretion column properties, here we use the model to simply obtain as accurate a description of the Her X-1 continuum emission as possible in order to robustly recover the broad Fe K line properties. The model has a number of physical parameters, some of which must be fixed. They are described in detail by the papers cited above. Below we describe the specific implementation of the model for our analysis.

In the model, we fix the neutron star radius and mass to 10 km and 1.4 M$_{\odot}$, respectively, and set the normalization to 1 \citep[as required by the model,][]{Ferrigno+09}. The mass accretion rate of Her X-1 is set to $1.6\times10^{17}$ g/s, which is obtained by taking the observed $1.8-75$ keV luminosity of $\sim3\times10^{37}$ erg/s, and calculating the mass accretion rate required to emit such luminosity \citep[assuming unitary accretion efficiency,][]{Wolff+16}. In principle, we could iterate with the model over a range of mass accretion rates until the output model luminosity precisely matches the mass accretion rate \citep[as shown by][]{Thalhammer+21} but this is unlikely to significantly change quality of the model fit to the broadband continuum \cite[see section 4.6 of][]{Wolff+16}. It would shift the best-fitting physical column parameters, but this is not of importance to this study which focuses on the broad Fe K emission line rather than on the interpretation of the accretion column emission. All of the above parameters are fixed and are not free variables of the spectral fit.

The variable parameters of the BW model are the accretion column radius $r_0$ (at the neutron star magnetic polar cap), the Comptonizing temperature of the radiating plasma $T_e$, the neutron star magnetic field $B$, and the dimensionless parameters $\xi$ and $\delta$, which describe photon diffusion and the relative contribution of bulk to thermal Comptonization, respectively \citep[for a detailed explanation, see][]{Ferrigno+09}. As the BW model assumes emission from the accretion column in its rest frame, following \citet{Thalhammer+21} we apply a \textsc{zashift} component in \textsc{xspec} to shift the model emission into observer frame. For a standard neutron star with a mass of 1.4 M$_{\odot}$ and a radius of 10 km, assuming that the accretion column emission originates at the polar cap (right above the surface), this redshift is roughly z$~\sim0.3$.

Another notable spectral feature of Her X-1 is its electron cyclotron resonance scattering feature at $\sim35$ keV \citep{Truemper+78}. This is prominently observed in the \nustar\ energy band \citep{Fuerst+13}. We model this feature with a \textsc{gabs} multiplicative Gaussian absorption model, and fit for the line energy, optical depth and line width. The line energy $E_{\rm{CRSF}}$ is related to the neutron star surface magnetic field (in the units of $10^{12}$ G) by the following equation: $B_{12} = (1+z)E_{\rm{CRSF}}/11.57$, where $z$ is the surface gravitational redshift ($z\sim0.3$). We link the best-fitting line energy in our spectral model to the magnetic field in the \textsc{bwcycl} component using this equation. The remaining parameters of the cyclotron feature (its optical depth and width) are free variables of the spectral fit. We do not model the first harmonic of the cyclotron scattering feature (expected around $\sim70$ keV) as it is very close to the edge of our energy band, where Her X-1 X-ray flux is beyond its peak, and so this simplification should not affect the recovered properties of the broad Fe K emission line. The final continuum emission model is therefore, in symbolic form, \textsc{gabs $\times$ zashift $\times$ bwcycl}.

Following our approach in \citet{Kosec+22} and more recently in \citet{Kosec+26}, we describe the complex Fe K energy band of Her X-1 using a combination of phenomenological models. This band contains a large number of emission and absorption lines in addition to the accretion column primary continuum emission. The highly broadened Fe K emission line, the main focus of this paper, is described with a Gaussian model, to obtain as model-independent a description of the line properties as possible, without any physical assumptions on its emission process or its emission region. We note that we do not include in any spectral fits the $\beta$ component (or any higher order transitions) of this emission line, because we do not know the atomic transition producing the line or the exact line emission process. Therefore, it is not clear what the centroid position of the $\beta$ line or its relative strength with respect to the $\alpha$ line should be.

Other spectral features in this energy band include a medium-width emission line consistent with the Fe XXV transition, and narrow emission lines at Fe I K$\alpha$ and K$\beta$ rest-frame energies. These 3 features are also fitted with Gaussian model components. As the Fe K$\beta$ feature is somewhat weaker, we couple its energy and width to the parameters of the Fe K$\alpha$ line, multiplied by a factor that is the ratio of the Fe K$\beta$ and K$\alpha$ rest-frame energies. We measure the luminosities of each Gaussian line using the \textsc{cglumin} component in \textsc{xspec}.

The highly ionized disk wind of Her X-1 \citep{Kosec+20} imprints narrow absorption lines on top of the Her X-1 emission continuum. We model these narrow features to correctly describe the entire X-ray spectrum, and not bias the broad Fe K emission line properties. In contrast to \citet{Kosec+26} where we used the \textsc{spex} fitting package, we are now fitting spectra in \textsc{xspec}, which does not natively contain ionized absorption models such as \textsc{slab} or \textsc{pion} that were used in our previous paper. We could in principle use the \textsc{xstar} photoionization model \citep{Kallman+01} to describe the disk wind absorption, but for a closer comparison with our previous results, we instead export the \textsc{slab} phenomenological ionized absorption model from \textsc{spex} into \textsc{xspec}. We use the exported \textsc{slab} component \citep{Kaastra+02} as a multiplicative table model in \textsc{xspec}, following the approach of \citet{Parker+09}. As opposed to that paper (see their Appendix A), which focused on exporting the \textsc{pion} model, \textsc{slab} outputs do not depend on the continuum spectral energy distribution of the source, and so we are able to export this model once as a table and apply this table to all of our pulse bin spectral fits. The exported \textsc{slab} table is a simplified version of the original model in \textsc{spex}, with only 3 free parameters: the velocity width of the absorption grid, and the column densities of the Fe XXV and XXVI transitions. These 3 parameters, in addition to the absorber blueshift free parameter (which is added to the multiplicative table directly in \textsc{xspec}) are sufficient to describe the most relevant ionized disk wind absorption features in the Fe K energy band of Her X-1, and result in a comparable spectral model as used in the phenomenological analysis part of \citet{Kosec+26}. We made a direct comparison between the results of the native and the exported \textsc{slab} model, and found that the results are consistent. 

Finally, all the components described above are obscured by interstellar absorption. To describe this effect, we use the \textsc{tbabs} model \citep{Wilms+00}, and set the neutral hydrogen column density value to $1\times10^{20}$ \pcm\ \citep{Kosec+22}.

To test how this model performs, we first apply it to time-averaged spectra from Orbits 1 to 3 in the following subsection. As a secondary check, to assess how this complex, partially physically-motivated, full-energy band model influences the derived Fe K emission line properties, we also performed a narrow-band spectral fit in the $2-12$ keV range. For this check we only used \xrism/Resolve Hp data and only applied phenomenological spectral models, using pulse-phase-resolved data from Orbit 2. The results of the comparison are shown in Appendix \ref{app:narrowband} and demonstrate that the line properties only weakly depend on the choice of the continuum model.

\subsection{Time-averaged spectra}

\begin{figure*}
\begin{center}
\includegraphics[width=\textwidth]{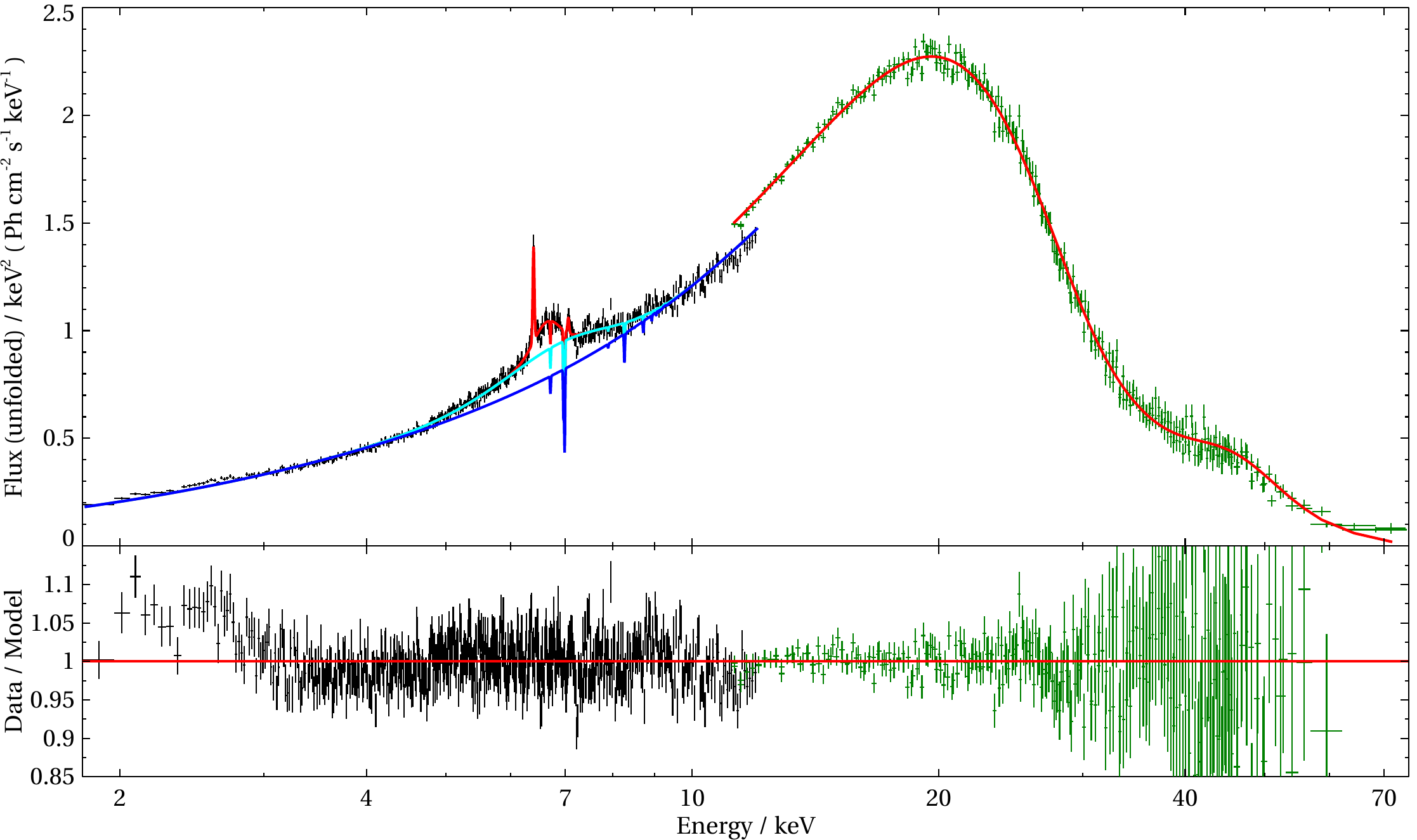}
\caption{Time-averaged \xrism\ (black) and \nustar\ (green) spectrum (top panel) and ratio residuals (lower panel) of Orbit 2, fitted with the spectral model described in section \ref{sec:modelsetup}. For visual purposes, only the Hp event \xrism\ data are shown and are heavily over-binned. Both FPMA and FPMB \nustar\ data are shown (superposed). The best-fitting model is split into components in the \xrism\ energy band: blue is the primary continuum (the \textsc{bwcycl} component), cyan shows the contribution of the broad Fe K line. The total model, which includes narrow Fe K$\alpha$ and K$\beta$ emission lines as well as a broadened Fe XXV emission line, is shown in red. All components include narrow line absorption from the ionized disk wind of Her X-1. The $\sim10$\% jump between the \xrism\ and the \nustar\ data is due to the usage of cross-calibration constants, which account for any residual differences between the instrument calibrations, and any real variations in the Her X-1 X-ray flux over time considering the \nustar\ exposure only covers part of the \xrism\ exposure. \label{Fig_orb2_stacked_SED}}
\end{center}
\end{figure*}

To determine the mean properties of the broad Fe K feature in Her X-1, we fit the time-averaged spectra from Orbits 1 to 3. First, we consider the data from Orbit 2. Orbit 2 is the highest S/N dataset because the Her X-1 flux was highest during that interval, and at the same time the \xrism\ exposure covered the entirety of that orbit. The best-fitting model, along with the residuals, is shown in Fig. \ref{Fig_orb2_stacked_SED}.

Overall, the model captures the time-averaged broadband spectrum of Her X-1 quite well, particularly in the 3 to 60 keV band where it matches the data to within $\sim5$\%. Larger differences are observed at both the very soft and the very hard end of the $1.8-75$ keV energy band. The positive excess of $5-10$\% below 3 keV could indicate the presence of an additional spectral component that is not currently modeled. The physical origin of this component may be highly broadened line emission from high-ionization Si and S lines near $2.0-2.5$ keV. However, such a component has not been previously observed in Her X-1. Alternatively, the excess could be due to instrumental residuals in the calibration of Resolve at the softest energies, which are most affected by the unopened gate valve. The observed excess is also similar to what would be expected from the electron-loss continuum \citep{Eckart+18} in \xrism/Resolve at low energies, but this effect is already accounted for in our analysis because we are using the XL-sized Resolve response matrices. We will investigate the possibility that this may be a new spectral component of Her X-1 in future work.

At the upper energy range above 60 keV, the model deviates from the data because we have not modeled the harmonic of the cyclotron resonance scattering feature, which should be located around 70 keV. Additionally, the source counts at $\sim70$ keV begin approaching the background level. Finally, there remains some curvature in the spectral residuals between 3 and 12 keV, but it only seems to be present at $\lesssim5$\% level.

The two main issues occur at the very edges of the modeled energy band and should not significantly affect our modeling of the Fe K energy band. Additionally, due to the strong spectral variability of Her X-1 over the pulse period, it may not be possible to completely describe the time-averaged spectrum with this relatively simple model. Therefore, we proceed with this model in our time-averaged as well as pulse-phase-resolved spectral analysis, but perform checks of our fitting approach where relevant.

\begin{figure}
\begin{center}
\includegraphics[width=\columnwidth]{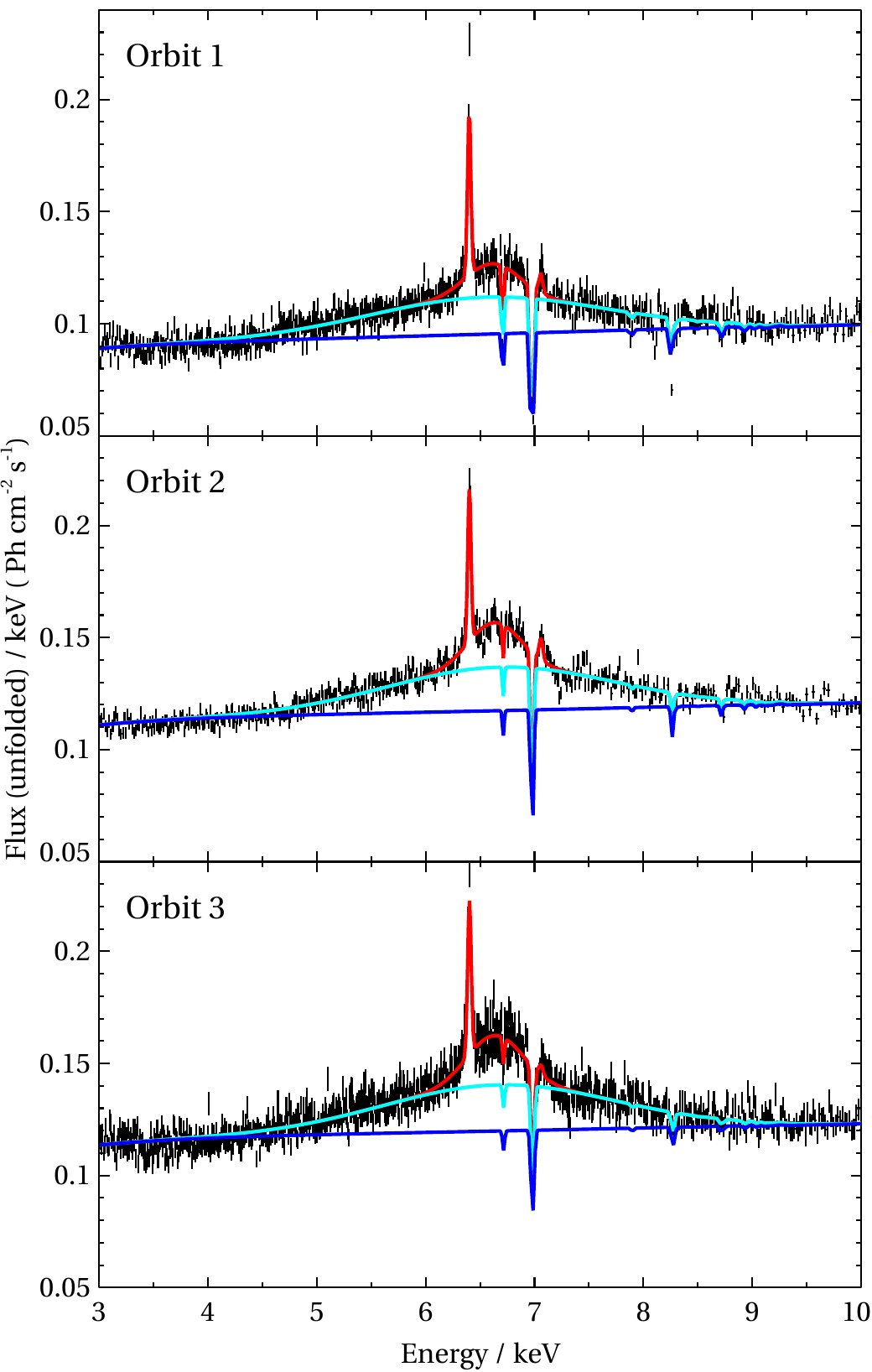}
\caption{Best-fitting time-averaged spectra from Orbit 1, 2 and 3, focusing on the Fe K region. Only Resolve Hp quality data are shown for visual purposes, and they are highly over-binned after fitting. Blue model is the primary continuum (the \textsc{bwcycl} component), cyan color shows the contribution of the broad Fe K line, and the total model (which includes narrow Fe K$\alpha$ and K$\beta$ emission lines as well as a broadened Fe XXV emission line) is shown in red. We note that the Y-axis units are different than in Fig. \ref{Fig_orb2_stacked_SED} to highlight the broad Fe K line. \label{Fig_time_av_zoom}}
\end{center}
\end{figure}

The same model is also applied to Orbits 1 and 3. The best-fitting results for the broad iron line from the time-averaged analysis are shown in Table \ref{res_time_averaged}. The best-fitting line energy is $6.4-6.5$ keV, and its typical $1\sigma$ width is $1.05-1.1$ keV, confirming previous results from the \xmm\ observations of Her X-1 \citep{Kosec+22}. Fig. \ref{Fig_time_av_zoom} shows a zoom-in on the Fe K energy band for all 3 datasets, showing the significance and smoothness of the broad Fe K line in the time-averaged, high sensitivity data. 

From Orbit 1 to Orbits 2-3, we observe a statistically significant change in the best-fitting broad Fe K line energy from 6.38 keV to $6.50-6.52$ keV. The reason for this jump is not known but one difference between these three datasets is that Orbit 1 does not have simultaneous \nustar\ coverage. To test that this difference did not introduce the observed jump in line energy, we performed a spectral fitting check using only the \xrism\ time-averaged datasets (i.e. excluding the NuSTAR data for orbits 2 \& 3), applying the phenomenological spectral model described in Appendix \ref{app:narrowband}. We fitted this simplified model to Hp quality time-averaged Resolve data from the three orbits (using only the $2-12$ keV energy range). In these simplified spectral fits, the best-fitting broad Fe K line energy for Orbits 1, 2 and 3 is $6.36 \pm 0.03$ keV, $6.50 \pm 0.02$ keV and $6.48 \pm 0.02$ keV, respectively. These values are consistent at $1\sigma$ significance with the results from the full spectral analysis and suggest that the lack of simultaneous \nustar\ data for Orbit 1 does introduce the observed jump in the time-averaged energy of the broad Fe K line.

In addition to the broad Fe K line, we also measured the properties of the broadened Fe XXV emission line and the narrow Fe K$\alpha$ and K$\beta$ lines during all 3 Her X-1 orbits. The complete spectral fitting results for all line components and the full BW model are listed in Appendix \ref{app:fullresults}. The typical time-averaged best-fitting energy of the Fe XXV line is $\sim6.62$ keV and its $1\sigma$ width is 0.25 keV. The typical width of the narrow K$\alpha$ and K$\beta$ lines is in the range of $15-20$ eV and their centroid positions are consistent with Fe I. A detailed analysis of these spectral components, which likely originate in different physical regions of the accretion flow than the broad Fe K line, will be presented elsewhere.

\begin{table*}
\caption{Results of the broad Fe K line modeling using time-averaged spectra from Orbits 1, 2 and 3. \label{res_time_averaged}}
\begin{center}
\begin{tabular}{ccccc}
\hline
\hline
Orbit&Line Luminosity& Equivalent Width& Line Energy&Line $1\sigma$ width\\
&$10^{35}$ erg/s&keV&keV&keV\\
\hline
1 & $ 3.17^{+0.05}_{-0.05} $ & $0.430^{+0.020}_{-0.013} $& $ 6.38^{+0.03}_{-0.03} $ & $ 1.09^{+0.02}_{-0.02} $\\
2 & $ 3.60^{+0.05}_{-0.05} $ & $0.395^{+0.012}_{-0.014}$ & $ 6.52^{+0.02}_{-0.02} $ & $ 1.05^{+0.02}_{-0.02} $\\
3 & $ 3.90^{+0.06}_{-0.06} $ & $0.413^{+0.012}_{-0.016}$ & $ 6.50^{+0.02}_{-0.02} $ & $ 1.07^{+0.02}_{-0.02} $\\
\hline
\end{tabular}
\end{center}
\end{table*}

\subsection{Orbit-resolved, pulse-phase-resolved analysis}

\begin{figure*}
\begin{center}
\includegraphics[width=0.55\textwidth]{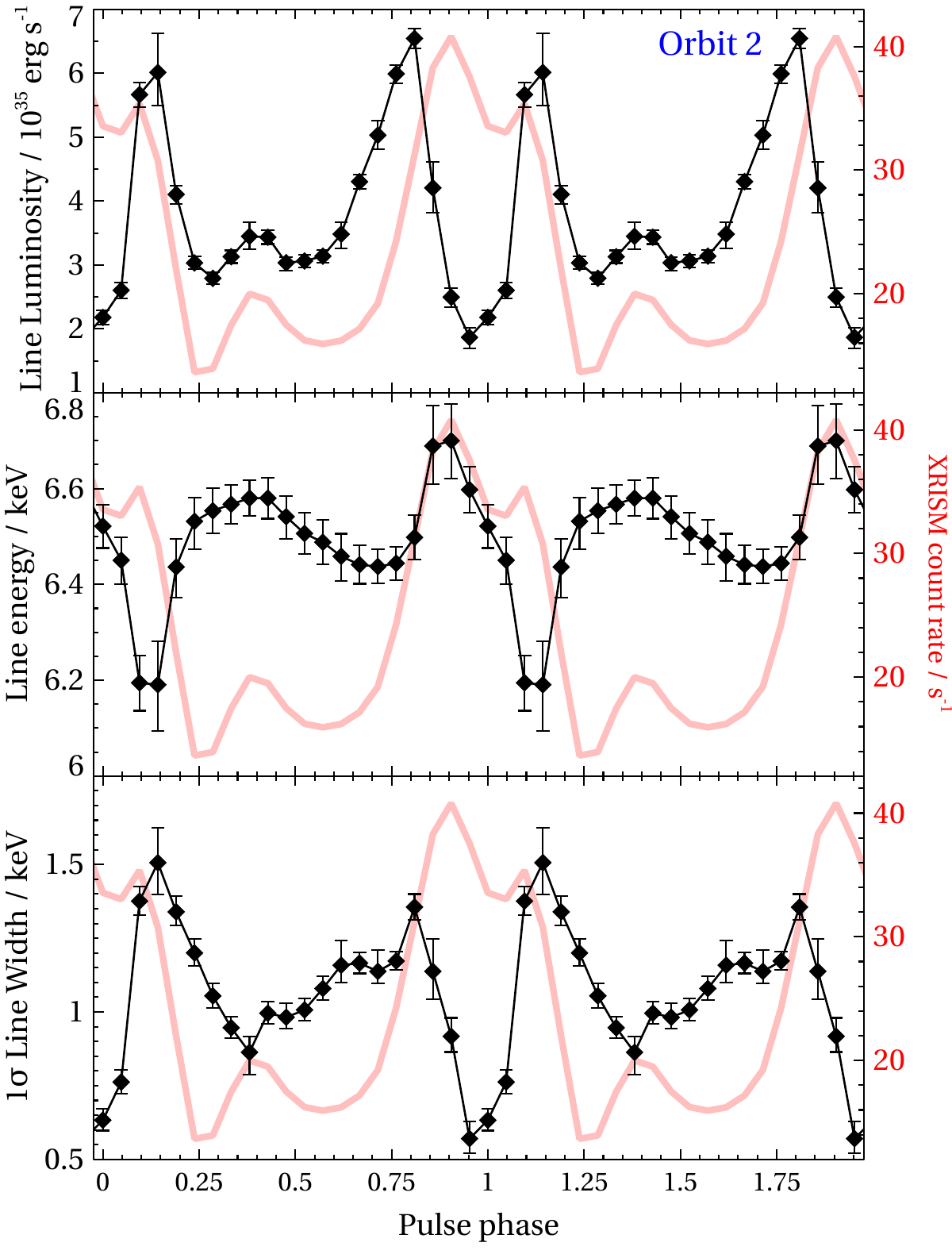}
\caption{Variation of the broad Fe K line with pulse phase of Her X-1 during Orbit 2. Two pulse periods are shown. The top panel shows the line luminosity, the middle panel the line energy and the lower panel the line width ($1\sigma$). For comparison with continuum pulsation, the \xrism/Resolve count rate is shown in red color in all panels. \label{Fig_orb2_results}}
\end{center}
\end{figure*}

We apply the model described in section \ref{sec:modelsetup} to the pulse-phase-resolved spectra of Her X-1. We again begin with Orbit 2 which offers the highest S/N and hence the individual pulse bins have the highest count statistics. Each pulse bin is fitted independently with the same baseline spectral model and we determine the best-fitting model parameters and their uncertainties. To speed up the fitting process, we employed the \textsc{parallel error} command in \textsc{xspec}, which uses multiple CPU cores and allows us to search all parameter errors simultaneously. We focus on the results for the broad Fe K emission line. The evolution of the Fe K line parameters with pulse phase is shown in Fig. \ref{Fig_orb2_results}.

We find that all the line parameters - flux/luminosity, energy and width - of the broad Fe K feature are strongly variable over the Her X-1 pulse period. The line luminosity does not mirror the continuum pulse profile evolution (also shown in Fig. \ref{Fig_orb2_results}), which shows a dominant main peak. Instead, the Fe K line luminosity variation shows two clear peaks which appear to surround the main peak of the continuum pulsation. The line luminosity varies by up to a factor of 3 during the pulse period. The line width evolves between 0.5 and 1.5 keV ($1\sigma$) and its variation is also double peaked, qualitatively following the behavior of the line luminosity. Finally, the line energy shows a different behavior. It varies between 6.2 and 6.7 keV and varies most strongly during the main peak of the continuum pulsation. First, it rises from 6.5 to 6.7 keV, followed by a brief decrease to 6.2 keV and a recovery back to 6.5 keV at the end of the continuum pulsation main peak.

\begin{figure*}
\begin{center}
\includegraphics[width=\textwidth]{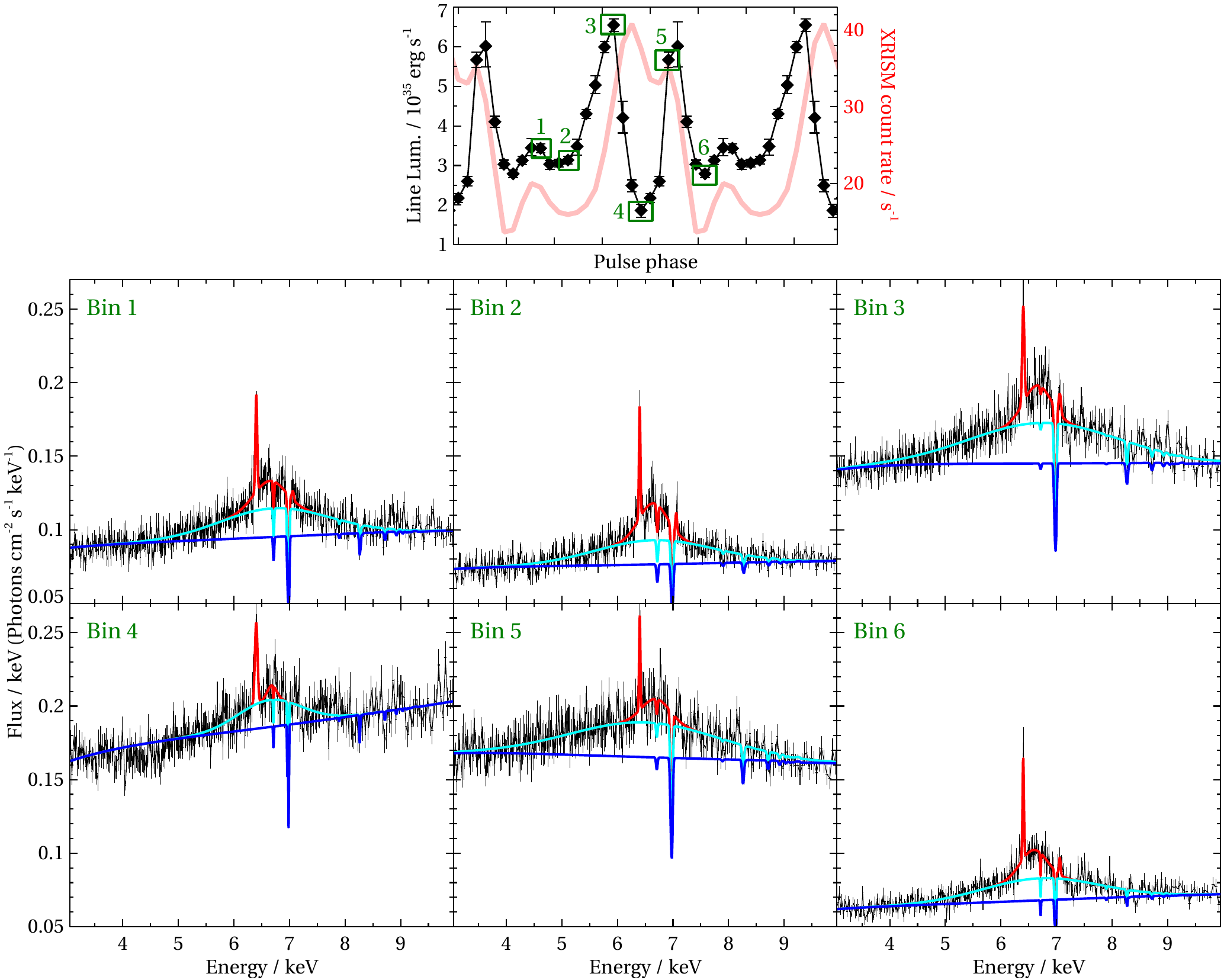}
\caption{Representative pulse bins from Orbit 2 illustrating the variations of the Fe K energy band of Her X-1 and the evolution of the broad Fe K line over the pulse period. The blue model shows the contribution of the primary continuum and cyan shows the broad Fe K line. Red represents the total spectral model, which includes narrow Fe K$\alpha$ and K$\beta$ emission lines as well as a broadened Fe XXV emission line. All components are modified by narrow line absorption by the highly ionized disk wind. The X and Y axis units are identical in all 6 panels, allowing direct visual comparison of the broad line properties.  \label{Fig_orb2_spectra}}
\end{center}
\end{figure*}

To visualize the variation of the broad Fe K line over the pulse period during Orbit 2, we show 6 different pulse bin spectra in Fig. \ref{Fig_orb2_spectra}, illustrating how the broad Fe K line evolves. We choose the most representative pulse bins over the pulse period, e.g. 2 bins close to the double peak maxima, 1 bin during the minimum between the two pulse peaks, as well as other bins during more monotonic parts of the cycle. The pulse bin spectra clearly show that the line strongly evolves in all its parameters. All individual pulse bin spectra were visually inspected and show consistent results.

We note that the narrower emission lines in the Fe K band, the Fe XXV and the Fe K$\alpha$ and K$\beta$ features, are also variable with pulse phase. The minimum line width of the broad Fe K line over the pulse period reaches $0.5-0.8$ keV, slowly approaching the typical $1\sigma$ width of the Fe XXV feature of 0.25 keV. We double checked our fitting results for these pulse bins to ensure that no interaction or switching occurs between the broad Fe K and the Fe XXV line. For all individual pulse bins, we found that the width of the Fe XXV line is at least a factor of 2 lower than that of the broad Fe K line, and therefore their joint fitting should not significantly affect or bias the determined best-fitting line parameters.

We also checked the influence of the observed $2-3$ keV residuals in \xrism/Resolve, which were discussed in the previous subsection. Especially if variable with pulse phase, these residuals could in principle systematically shift our underlying broadband continuum description for certain pulse bins, which would bias our fitting results for the broad Fe K line. To check that our fitting results are robust to these residuals, we excluded all \xrism/Resolve data below 3 keV and repeated both the full-band (Section \ref{sec:modelsetup}) and the narrow-band (Appendix \ref{app:narrowband}) spectral analyses. In both cases, the broad line is still observed to be highly variable in all its parameters, and the variation pattern is consistent with the results presented in Fig. \ref{Fig_orb2_results}. Specifically, in both these modified analyses we observed the same double-peaked pattern and the drop between the peaks in the line luminosity and width evolution, as well as the single sudden decrease in the centroid energy.

\begin{figure*}
\begin{center}
\includegraphics[width=\textwidth]{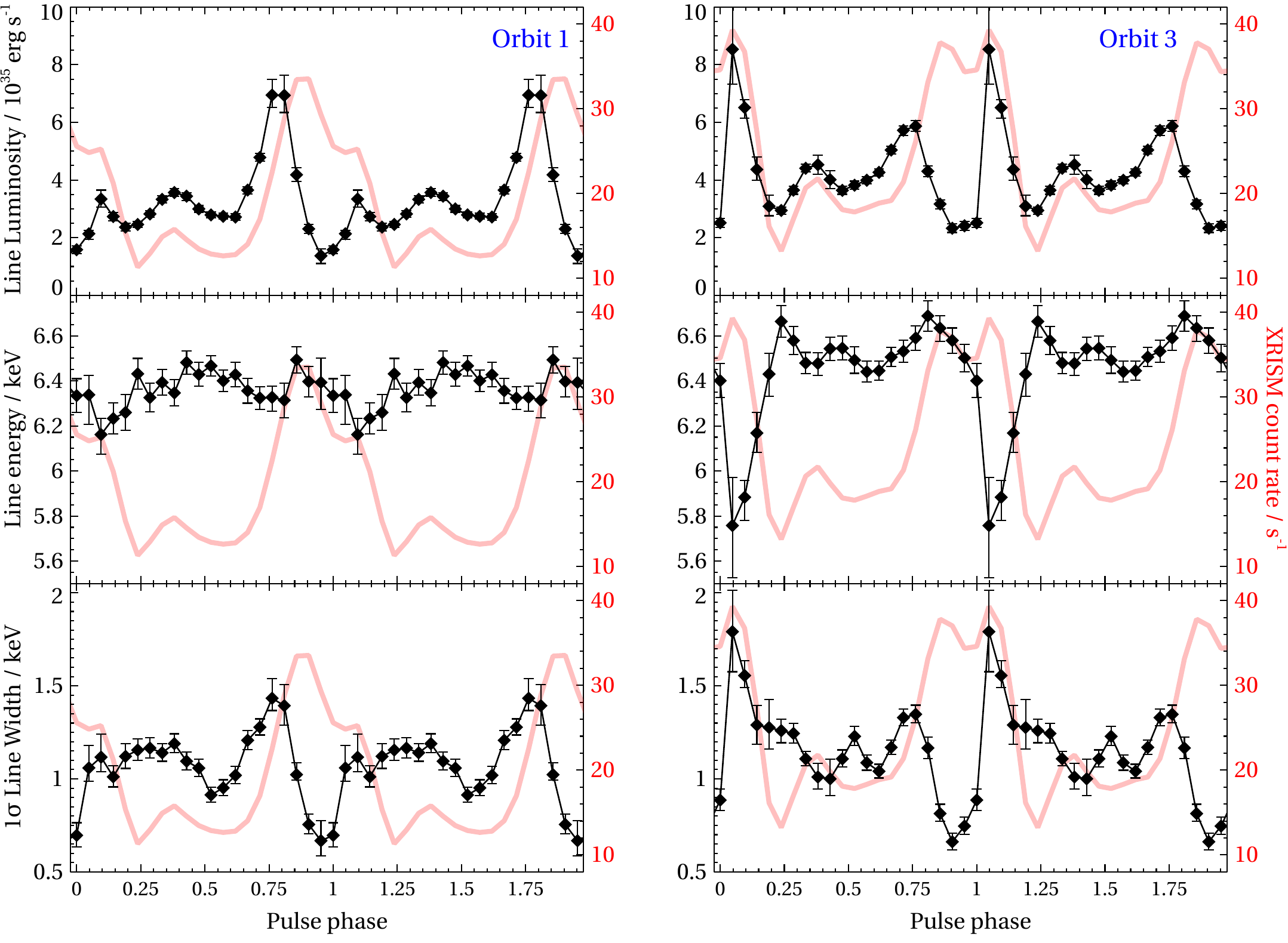}
\caption{Variation of the broad Fe K line with pulse phase of Her X-1 during Orbit 1 (left panels) and Orbit 3 (right panels). Two pulse periods are shown. The top panel shows the line luminosity, the middle panel the line energy and the lower panel the line width ($1\sigma$). For comparison with continuum pulsation, the \xrism/Resolve count rate is shown in red color in all panels. \label{Fig_Orb1_Orb3}}
\end{center}
\end{figure*}

Secondly, we apply the same analysis to Orbits 1 and 3. The best-fitting results for these orbits are shown in Fig. \ref{Fig_Orb1_Orb3}. Nearly all line parameters are again strongly variable over the pulse period, but the behavior of the broad Fe K line during these orbits is qualitatively different. We do not see a clear double-peaked luminosity evolution in Orbit 1. Instead there is one strong peak, which precedes the main peak of the continuum pulsation. The line width evolution appears to follow this behavior. In Orbit 3, there is again evidence for two peaks in the line luminosity evolution, but the second one is stronger, as opposed to Orbit 2 where neither was dominant. Again, the line width follows the line luminosity behavior. The line energy evolves only weakly during Orbit 1, while it shows similar behavior in Orbit 3 to the behavior seen in Orbit 2. Orbit 3 also shows the most extreme change of the line energy, when the line centroid drops down to just 5.8 keV, as opposed to a typical centroid energy of $6.2-6.7$ keV.

\section{Discussion}
\label{sec:discussion}

We performed a time-averaged and pulse-phase-resolved analysis of the 200 ks \xrism\ observation of Her X-1 taken in September 2024, focusing on the properties of the broad Fe K emission line found previously using lower-spectral resolution instruments \citep{Fuerst+13, Asami+14, Kosec+22}. The line is detected at high statistical significance, and now for the first time, thanks to \xrism/Resolve, it is fully resolved from other complex spectral components in the Fe K band, which include narrower emission lines \citep[originating in various locations of the accretion flow,][]{Kosec+22} as well as narrow absorption lines from the ionized disk wind \citep{Kosec+20, Kosec+26}.

This is the first attempt to describe this spectral feature and its shape using high-spectral resolution data. We used phenomenological modeling to begin understanding how the line evolves over the pulse period of Her X-1, as well as how it changes with increasing precession phase of the 35-day cycle of Her X-1, without making strong assumptions on the line emission region or specific emission process.

\subsection{Broad Fe K line origin and evolution with pulse phase}

We found that the feature is strongly variable in all its parameters over the pulse period. The superior resolution of \xrism/Resolve shows that the line is indeed extremely broad and smooth (confirming the results of previous lower-resolution observations) and reaches $1\sigma$ widths of 0.5 to 1.5 keV. Additionally, the line width variation is very rapid and we observe variations across this entire range within 15\% of the pulse period, i.e. within 0.2 seconds of real time. Similarly, we observe rapid and significant variations in the line flux and centroid energy. The variation patterns of line luminosity and 1$\sigma$ width are relatively similar, but the variation of the centroid energy is very different. Furthermore, the variation patterns of these quantities evolve over the three Her X-1 orbits studied in this work.

Assuming the broadening is due to Doppler shift, the measured line width (Fig. \ref{Fig_orb2_results} and \ref{Fig_Orb1_Orb3}) indicates velocities from 0.08c to 0.25c along our line of sight. Such high velocities are only present in the magnetosphere of the neutron star, as the disk is truncated at $\sim1000$ R$_{\rm G}$ and hence the maximum Keplerian velocity reaches just $\sim0.03$c. The line emission should therefore originate somewhere in the magnetosphere, in the infalling material which follows the neutron star magnetic field lines and strongly decelerates in the accretion column. The observation of Doppler velocities in excess of 0.2c at certain portions of the pulse cycle additionally indicates that the emitting material is located in the inner part of the magnetosphere. Moreover, given the expected velocity projection effects (as the material likely does not move directly along our line of sight), the 3D velocity of the emitting material is most likely even higher than the largest observed line width. Thus, at least some fraction of the emitting material may actually be very close to the neutron star surface ($\lesssim25$ R$_{G}$ from the star), where velocities as high as $0.3-0.5$c are present \citep{Basko+76, Becker+98, Becker+22}. The observed strong variations in the line parameters are then naturally explained by the motion of the accretion stream and column as they rotate alongside the neutron star magnetosphere.

We observe a strong variation of the line luminosity with pulse phase that does not mirror the pulsation of the primary continuum. In Orbit 2, the luminosity pulsation is a double-peaked function, surrounding the main continuum pulsation peak. Additionally, in all 3 orbits, a strong rise in the line flux occurs before the primary continuum begins to rise into the main peak. Therefore, the emission regions of the Fe K line must be sufficiently distinct from where the primary X-ray continuum is emitted in the accretion column, to produce this completely different pulsation pattern. The line could originate in the reprocessing of the primary X-ray emission off of the column itself (at greater distances away from the neutron star, Fig. \ref{Pe_HerX1_scheme}). Alternatively, it could be produced by collisional recombination in a similar region of the accretion flow. The fact that we observe the line luminosity to rise first suggests that as the neutron star magnetosphere rotates, the upper portion of the column or stream (where the Fe K emission originates) comes into our view first, followed by the lower part of the column which emits the primary continuum. We note that any phase delays between the continuum and broad Fe K line flux cannot be due to light travel time delays. Considering the very large line width, much of the line flux is likely emitted within 100 R$_{G}$ ($\sim200$ km) from the neutron star. Any light travel time delays between the continuum and line flux are therefore most likely smaller than $10^{-3}$ s. The observed phase delays between the line and continuum flux can instead originate in time-variable obscuration of the accretion columns, or due to relativistic boosting and light bending effects.

As the line most likely originates in the vicinity of the neutron star, other phenomena could contribute to the line width besides the emitting region bulk velocity field. Turbulent motions of the sinking material, decelerated from the free-fall velocity in the accretion column may broaden any Fe K emission, if originating in this part of the column. Turbulence would produce a Gaussian line shape in agreement with the observed line shape. However, it is not clear if it could achieve a velocity broadening in excess of 0.2c, and introduce the observed fast variations in all the line properties including its centroid energy. We note that even in this case, we would be observing a direct observational signature of the accretion column.

Additional line broadening may be provided by gravitational redshift, which reaches $z\sim0.3$ in the vicinity of the star. The line cannot originate on the surface of the neutron star as the surface is nearly stationary (given the long rotation period of Her X-1) and so could not produce such a broad line, only a highly redshifted one. However, if the line originates in the accretion column or stream over a range of heights above the neutron star, the varying gravitational redshift will broaden the observed Fe line beyond just the underlying Doppler motions. This effect would likely produce a line skewed redward as the rotation of the magnetosphere would average out the bulk velocities in the column, but gravitational redshift will always affect the line energy in the same direction.

A further effect that may shift the line properties in the very vicinity of the neutron star is that of the strong magnetic field. Strong fields will will shift the Fe transition energies beyond the laboratory rest-frames, as their ionization potential increases \citep{Ruder+94}. This effect will as a result increase the observed line energy, but is only of importance if the line originates within a few radii of the neutron star. Evidence for this effect was recently found in the neutron star RX J0822-4300 \citep{Groger+26}.

Alternatively, a broader line could be achieved by Comptonization of a narrower line profile, which was previously considered by \citet{Asami+14}. The $1\sigma$ width of a Comptonized line profile can be derived as follows:

\begin{equation}
    \sigma=E\left[\frac{2kT}{m_{e}c^{2}}+\frac{2}{5}\left(\frac{E}{m_{e}c^{2}}\right)^{2}\right]^{\frac{1}{2}}
\end{equation}

where $E$ is the line energy, $k$ is the Boltzmann's constant, $T$ is the gas temperature and $m_e$ is electron mass \citep{Ross+78, Garcia+13}. To achieve the observed time-averaged line width of 1.07 keV (Table \ref{res_time_averaged}), the necessary plasma temperature is 7 keV, but to achieve the maximum line width of $\sim1.8$ keV (observed during Orbit 3), a temperature of 20 keV is needed. The Her X-1 system exhibits an accretion disk corona in the outer part of the accretion flow with a temperature of $200-400$ eV \citep{Jimenez+05, Ji+09} which could Compton scatter photons from the inner accretion flow, but its temperature is far too low to provide all of the needed Fe K line width. Another possibility is Compton scattering in the accretion disk wind, but the temperature of the observed wind plasma is still insufficient at $T\sim1$ keV \citep[calculated using the best-fitting wind parameters from][]{Kosec+26}. Even the emitting surface of the accretion column only reaches temperatures of $\sim5$ keV \citep{Wolff+16}. It is not clear if an even hotter scattering corona can exist in Her X-1. There is evidence for a large-scale scattering corona observed during the eclipses of Her X-1 \citep{Leahy+15}, which may be related to the disk wind or the disk corona, but its temperature is unknown.

Considering the very fast changes of the line width and particularly its centroid with respect to the pulse phase, our interpretation is that Doppler shift is the dominant mechanism providing the line width. Other mechanisms as discussed above, such as gravitational redshift, could contribute to the observed line broadening, but are unlikely to dominate it. 

In our phenomenological analysis, we found that the broad Fe K line can be well reproduced to first order with a simple Gaussian component. This suggests that the line is not strongly asymmetric (e.g. as observed in relativistic Fe K reflection in accreting black holes), as is visually evident in Fig. \ref{Fig_time_av_zoom} and \ref{Fig_orb2_spectra}. However, a lower degree of asymmetry, introduced for example by a varying gravitational redshift across the line emission regions, may still be present, and would require more complex spectral models to quantify. We will explore this in follow-up work.

The observed smooth, simple line shape is relatively puzzling considering that two accretion streams and columns are likely present in Her X-1. In that case, we may naively expect a double-peaked line shape, which is not seen. However, it is possible that at least at most phases of the pulse cycle, one of the accretion columns is oriented such that its Fe emission is blocked from our line of sight by either the neutron star itself, or the other column or accretion stream. Obscuration of the Fe line emission sites by the accretion stream would naturally explain the very fast jumps in the line luminosity and width over the pulse cycle. Secondly, as the line likely originates very close to the neutron star, relativistic light bending and boosting may also play an important role by increasing the flux of one of the two Fe K emission sites with respect to the other at certain portions of the pulse cycle. In this study, we have not explicitly attempted to fit the \xrism\ and \nustar\ data with two broad Gaussian lines, as doing so would further increase the degeneracy of an already complex spectral model, and the single Gaussian model provides a reasonably good spectral description.

Evidence for an eclipse of the primary accretion column emission by the neutron star was recently found in the transient X-ray pulsar V 0332+53 \citep{Mushtukov+24}. A similar type of phenomenon may be at play here in Her X-1 for the Fe K emission sites. However, we note that the primary X-ray continuum emission sites (near the base of the column) cannot be obscured at the same time as the Fe K emission sites, because the phase of the minimum of Fe K emission coincides with the maximum of the X-ray continuum pulsation.

A final, less likely possibility is that an accretion disk exists within the neutron star magnetosphere, and reprocesses the primary emission from the accretion column. In such a case, we may expect a typical relativistic line shape as observed in accreting black holes \citep[e.g.][]{Fabian+89} - a strong peak blueward of the transition rest-frame, and an extended red wing. This is not seen in our case, neither in the time-averaged data, nor in the individual pulse bins. However, we note that the observed relatively symmetric line shape can still be reproduced with a relativistic reflection disk model. We attempted to fit the time-averaged and pulse-phase-resolved Her X-1 spectra with the \textsc{relxill} model \citep{Garcia+14} version 2.8, and managed to achieve spectral fit quality comparable with our phenomenological Gaussian line analysis. The best-quality fits were achieved using the \textsc{relxillns} flavor of the model \citep{Garcia+22}, which is typically used to describe relativistic lines in low-magnetic field neutron stars. The best-fitting inner disk radii were in the range of $10-20$ R$_{\rm G}$ (in agreement with large observed line widths), and the disk inclination varied strongly over the pulse period. The significant variation in the best-fitting inclination further supports the line origin in the accretion column, as rapid changes in the reflection angle are naturally expected due to the rotation of the accretion column and stream. At the same time, we prefer not to over-interpret these results and do not proceed with further \textsc{relxill} spectral modeling at this stage as it may give us biased results given the more likely Fe line origin in the accretion column, where the Fe line emitting region has a very different geometry than the typical accretion disk case assumed in \textsc{relxill}.

In general, the line width variation with pulse phase appears to be similar to the evolution of the line luminosity. On the other hand, the line energy variation follows a very different pattern. The best-fitting energy of the line oscillates between 5.8 and 6.7 keV (the largest variations in the line energy are observed in Orbit 3). As the line is extremely broad and we only model this one specific feature using a phenomenological spectral model, it is challenging to identify the feature with a particular Fe K transition. The line energy varies outside of the range of the main Fe K transitions which occur between 6.4 keV (neutral or low-ionization Fe) and 6.96 keV (Fe XXVI). However, given the very high involved velocities of the emitting plasma and possible gravitational redshift, any of the Fe K transitions are plausible. Even if the emitting transition is Fe XXVI at 6.97 keV, gravitational redshift (which may be up to z$\sim0.3$) could alone explain the full range of the observed broad Fe K centroids.

If the line is due to reprocessing of the primary accretion column continuum, another important consideration is the ionization state of the emitting material, usually described by the ionization parameter $\xi=L/(nR^{2})$ \citep{Tarter+69}, where $L$ the ionization luminosity, $n$ its number density and $R$ its distance from the ionizing source (base of the accretion column). As the material progresses through the magnetosphere, if its density is low enough, it will be over-ionized and no lines should be observable. As the material follows the accretion column, if its density strongly increases, its ionization parameter may decrease. As a result, at some point the highest ionization transition, i.e. Fe XXVI may increase in strength, become detectable and could be observed as the broad Fe K line seen here. In principle, as the density keeps increasing, lower transitions (Fe XXV, XXIV) may become more prominent. In that case, the observed broad line could be a blend of multiple Fe K transitions. However, we caution that other interpretations may be possible. If the infalling material remains dense across the entire infall through the magnetosphere, low-ionization Fe K lines (with rest-frame centroids close to 6.4 keV) may be the dominant transitions at all times. Physically-motivated density calculations of the accretion stream and column are required to determine the expected ionization levels.

\subsection{Variation in the broad Fe K line pulsation pattern and potential connection to super-orbital precession}

\begin{figure*}
\begin{center}
\includegraphics[width=0.55\textwidth]{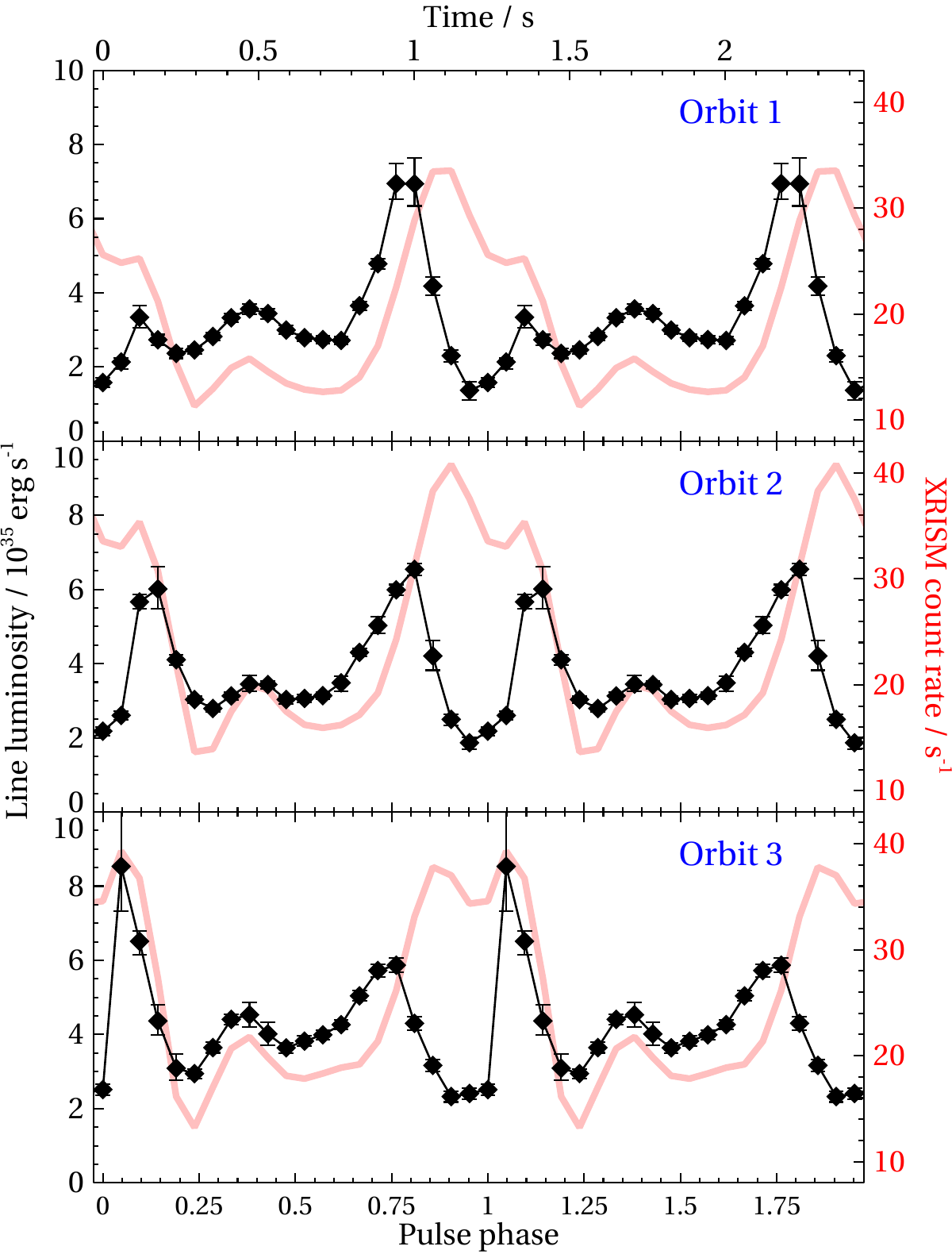}
\caption{Comparison of the broad Fe K line pulse pattern during Orbit 1 (top panel), Orbit 2 (middle panel) and Orbit 3 (lower panel). The panels show the line luminosity (black) and the \xrism\ count rate (red) versus pulse phase. \label{Fig_orb_combined}}
\end{center}
\end{figure*}

We observe a clear variation in the line pulsation pattern from Orbit 1 to Orbit 2 to Orbit 3. For a clearer comparison, we show the line luminosity variation with pulse phase for the three orbits in Fig. \ref{Fig_orb_combined}. During Orbit 1, the line pulsation pattern shows a single prominent peak which precedes the continuum pulsation. During Orbit 2, two clear peaks are seen to `surround' the continuum pulsation peak and have comparable strengths. Finally, two peaks are also observed during Orbit 3, but the second one appears to be stronger than the first one. From Fig. \ref{Fig_orb_combined}, it is further evident that the continuum pulsation peak also slightly evolves from orbit to orbit. The main peak appears to be composed of two smaller peaks, separated by $\sim0.25$ of pulse period. During Orbit 1, the first peak clearly dominates, but its strength decreases with increasing orbits, while the second peak increases in strength until they both roughly match in strengths in Orbit 3. This evolution appears to go hand-in-hand with the broad Fe K line pulsation pattern evolution, but the variations in the continuum variation are smaller than those of the broad Fe K line. We note that the Fe line luminosity pulsation pattern during Orbit 2 particularly resembles the double-peaked emission pattern of an accretion column fan beam as modeled for Her X-1 by \citet{Scott+00} and \citet{Blum+00}. It is thus possible that the emission of the broad Fe K line is related to the emission of the fan beam.

These orbit-to-orbit changes correspond to a changing precession phase of the 35-day cycle of Her X-1. This cycle of high and low X-ray flux states has been studied for the past 50 years \citep{Katz+73, Gerend+76}. The observed variations in the X-ray flux are explained if the accretion disk of Her X-1 is warped, and precesses every 35 days and periodically obscures the inner accretion flow, thus introducing the extended low states. However, the underlying cause of the warped disk precession is still not conclusively explained. Plausible interpretations include torquing of the disk due to the launching of a thermal wind in the outer disk \citep{Schandl+94} or radiation-driven warping \citep{Ogilvie+01}. However, it is possible that in addition to the warped disk precession, the neutron star itself also undergoes precession \citep{Staubert+09a} with a similar precession period to that of the warped disk, which is observed through a periodically varying continuum pulse shape. Strong evidence for neutron star precession in Her X-1 was recently also found by the \ixpe\ observatory, which detected a changing polarization angle of the Her X-1 X-ray continuum emission over time \citep{Doroshenko+22}. The neutron star and the warped disk precession cycles could be locked together, the former introducing the observed variable continuum pulse profile of Her X-1, and the latter introducing the observed high and low flux states.

Under this `double precession' scenario, the observed changing pulsation pattern of the broad Fe K line can potentially be explained. If the line originates in the accretion column close to the neutron star, and the star precesses with a 35-day cycle, the accretion column position and orientation with respect to our line of sight will change over time. Therefore, the projected velocities of infalling matter along the column and their positions will change as the magnetic pole precesses, resulting in a change in the pulsation pattern of the emitted broad Fe K line. The fact that the continuum pulsation shape appears to evolve in a similar matter to the broad Fe K line pulsation shape further supports this interpretation.

\subsection{Comparison with other emission lines in Her X-1}

The broad Fe K line is just one of several emission lines observed in the Fe K band of Her X-1. A narrow component, consistent with neutral or low ionization Fe transitions is present at 6.4 keV and has a velocity width of $\sim1000$ km/s \citep{Kosec+22}. Given its very low width, it likely originates in the outer part of the accretion flow, and so it is unlikely to be related to the broad Fe K line. However, a second, broader line component is present, with a $1\sigma$ width of 0.25 keV that is close to the rest-frame energy of the Fe XXV transition at 6.67 keV. As previously discussed by \citet{Kosec+22}, the velocity width of this line ($v\sim0.04$c) is consistent with the outer edge of the magnetosphere (and the inner disk) of Her X-1. Therefore, it is plausible that the Fe XXV line is emitted in the outer part of the magnetosphere, while the broad Fe K line originates from the inner part of the accretion column and as such their emission and variability patterns may be related. The properties of the narrow and the Fe XXV components will be discussed in Narita et al. (in prep.).

Another spectral feature of Her X-1 which may be of relevance is the 1 keV `bump'. A strong excess is observed between 0.8 and 1.2 keV and was first described by \citet{Jimenez+02} and analyzed in further depth by \citet{Kosec+22}. This residual can be described with a broad Gaussian line with a centroid energy of around $0.9-1.0$ keV, and a $1\sigma$ width of $0.13-0.2$ keV. If originating from a single line transition, the observed line width would indicate Doppler motions of $0.1-0.2$c, similar to the broad Fe K line, possibly suggesting a common origin. However, it is also possible that the line is a blend of many different Fe L transitions, with contribution from Ne lines \citep{Chakraborty+24}, reducing the required line velocity width. Nevertheless, a pulse-phase-resolved analysis of the 1 keV feature and its comparison with the broad Fe K line would be of great interest, which could reveal similarities or differences in variability and pulsation trends. We note that the time-averaged luminosity of the Fe L line is strongly correlated with that of the broad Fe K line \citep{Kosec+22}. Some pulsation of the Fe L complex is very likely considering the $0.8-1.2$ keV energy band of Her X-1 is strongly pulsed with a complex pulsation pattern \citep[see Fig. 2 of][]{Ramsay+02}.

\subsection{Comparison with other X-ray binaries}

It would be of great importance to confirm the same phenomenon of highly broadened, pulsating Fe K emission from the accretion column in other X-ray pulsars. Other X-ray pulsars are known to show pulsating Fe K lines, including Cen X-3 \citep{Roy+25} and V 0332+53 \citep{Bykov+21}, but the observed line widths are much lower ($<0.5$ keV), indicating that the lines may not originate in the accretion column. The best candidate to date is the transient ultraluminous X-ray pulsar Swift J0243.6+6124, which reached an X-ray luminosity of $\sim10^{39}$ erg/s during its 2017-2018 outburst \citep{Wilson-Hodge+18}. This source showed both a very broad Fe K emission line \cite[reaching a $1\sigma$ width of almost 1 keV,][]{Jaisawal+19} which is also pulsed \citep{Xiao+24}. Interestingly, the line in Swift J0243.6+6124 appears to be strongly asymmetric with a strong red wing \citep{Jaisawal+19}, resembling standard relativistic X-ray reflection, as opposed to the more symmetric line of Her X-1. However, at the same time there is strong evidence for a high magnetic field of $1.6\times10^{13}$ G in this system \citep{Kong+22}, indicating significant truncation of the accretion disk and thus again suggesting an accretion column origin for the line. A broad Fe K emission line with a $1\sigma$ width of up to 0.8 keV was also observed in the Be X-ray pulsar LS V +44 17 using \nicer\ \citep{Mandal+23}.

An important complication is that most X-ray pulsars have only been observed with moderate spectral resolution instruments in the Fe K band such as \nicer, \xmm\ or \nustar. Our study of Her X-1 with \xrism/Resolve demonstrated how complex the Fe K energy band can be in X-ray pulsars, and can contain multiple emission as well as absorption components.

We finally note that broad Fe K emission lines are also observed in accreting neutron stars with low magnetic fields \citep[B$<10^{10}$ G, e.g.][]{Ludlam+24}. In those systems, the broad line originates from the inner parts of the accretion disk, which is not highly truncated as the neutron star magnetic field is many orders lower than in X-ray pulsars. Therefore, the broad line resembles the typical accreting black hole relativistic reflection shape, and is clearly asymmetric in shape, as shown in the recent observations of Ser X-1 at high spectral resolution with \xrism\ \citep{Ludlam+25}.

\subsection{Future work and applications}

This is the first attempt to describe the observed broad Fe K line using high-resolution data from \xrism, with which we can easily separate this feature from other spectral components observed in the Fe K energy band. By describing the line with a simple Gaussian, we are able to observe how the feature changes with Her X-1 pulse and precession cycle in the most model-independent way.

To fully leverage this feature and learn more about the physics of the accretion column and the neutron star itself, a physical model will be required. We note that a standard relativistic reflection model as typically applied for accreting black holes \citep[e.g. \textsc{relxill},][]{Garcia+14} will not be sufficient to obtain accurate results considering the very different accretion geometry. In the case of the black hole accretion disk, a symmetric disk is assumed, and the velocity vector of the infalling material is perpendicular to the position vector from the black hole. Here, the infalling matter directly follows the (rotating) magnetic field lines, and its distribution is strongly non-axisymmetric. The physical model to describe the broad Fe K line will need to include relativistic beaming and light bending by the neutron star, as well as the obscuration effects of the accretion column by the star itself. Finally, the model may need to take into account the strong magnetic field effects on the atomic transitions \citep{Ruder+94}.

A physical model of the broad Fe K line will give us a powerful new tool to study highly magnetized neutron stars. The line width, centroid energy and flux/luminosity encode the accretion column properties such as its orientation and location at each phase of the pulse cycle. With modeling over the entire rotation cycle, these constraints may be sufficient to determine the orientation of the rotation and the magnetic axis of the neutron star. By sampling these parameters over time such as in the case of Her X-1 in this work, we may be able to probe shifts in these orientations over time and study neutron star precession. The resolution of \xrism\ is critical for this science, as only Resolve can constrain the broad line and accurately measure its shape in the presence of other spectral components.

Tracking free precession in neutron stars is a powerful tool to constrain the properties of matter at supra-nuclear densities in the neutron star interiors \citep{Link+07,Heyl+24}. This Fe K line tracking method naturally offers excellent synergies with \ixpe, which is able to probe similar physical quantities and phenomena in X-ray pulsars via X-ray polarization. A coordinated observational campaign with \xrism\ and \ixpe\ on an X-ray pulsar showing the broad Fe K line such as Her X-1 will be able to confirm and leverage this synergy.

\section{Conclusions}
\label{sec:conclusions}

We performed a detailed time-averaged and pulse-phase-resolved analysis of \xrism\ and \nustar\ observations of Hercules X-1 taken during the Main High state in September 2024. We specifically focused on the properties of the highly broadened Fe K emission line, which was described using phenomenological spectral models. Our main conclusions are as follows:

\begin{itemize}

    \item The high-spectral resolution of \xrism/Resolve allows us to separate the different spectral components in the Fe K band of Her X-1. We confirm the presence of a highly broadened, smooth emission line with a typical $1\sigma$ width of $1.0-1.1$ keV in the time-averaged Her X-1 spectra during all 3 Main High state orbits covered by our \xrism\ observation.

    \item Our pulse-phase-resolved analysis shows that the broad Fe K line is strongly variable over the pulse period in all its parameters. Its luminosity varies in a double-peaked pattern. The line energy varies between 5.8 and 6.7 keV, and its width between 0.5 and 1.8 keV. The variations are very rapid and occur on timescales as fast as 15\% of the pulse period, which is 0.2 s of real time.

    \item The most likely scenario is that the broad Fe K line originates in collisional recombination in the accretion column of the neutron star, or by reprocessing of the primary column emission off of the column itself. The high line broadening is naturally explained by the large free-fall velocities (up to $\sim0.5$c) present in the column. The rapid variation with pulse phase is introduced because the column rotates alongside the neutron star and its magnetosphere.

    \item  We analyse 3 consecutive Her X-1 orbits covered by these observations and show that the broad Fe K line pulsation evolves over time, with increasing phase of the 35-day precession cycle of Her X-1. If the origin of the line is in the accretion column, this suggests a varying accretion column orientation over time, in agreement with observed continuum pulse profile changes over the same period of time. These results indicate that the neutron star is precessing and are consistent with the recent \ixpe\ observations of Her X-1, which also revealed precession of the neutron star.

     \item This was the first attempt to describe the broad Fe K line in high-resolution X-ray spectra, where we used only phenomenological models to describe this feature. Future physically-motivated spectral models will put direct constraints on the accretion column location and orientation over the pulse period as well as over longer timescales. They have great potential to help us detect and track precession in neutron stars, and leverage it to constrain the physics of neutron star interiors.

\end{itemize}

\begin{acknowledgments}

PK thanks Petra Palenikova for help with the schematic of Her X-1 and Ralf Ballhausen for insightful and helpful discussions about Her X-1. Support for this work was provided by NASA through the NASA Hubble Fellowship grant HST-HF2-51534.001-A awarded by the Space Telescope Science Institute, which is operated by the Association of Universities for Research in Astronomy, Incorporated, under NASA contract NAS5-26555. PK also acknowledges support from NASA grants 80NSSC25K7317 and 80NSSC25K7533. CP is supported by European Union - Next Generation EU, Mission 4 Component 1 CUP C53D23001330006. DJW acknowledges support from the Science and Technology Facilities Council (STFC; grant code ST/Y001060/1). FB acknowledges support from INAF Large Grant 2023 BLOSSOM F.O. 1.05.23.01.13.
\end{acknowledgments}





%
\facilities{\xrism, \nustar}

\software{Veusz, SPEX \citep{Kaastra+96}, XSPEC \citep{Arnaud+96}
          }


\appendix

\section{Full spectral fitting results of the time-averaged analysis of Orbit 1, 2 and 3}
\label{app:fullresults}

This appendix contains Table \ref{res_timeavg_full}, which summarizes the results of our time-averaged spectral fitting analysis for all orbits.

\begin{table*}
\caption{Results of spectral fitting of the time-averaged data from Orbit 1, 2 and 3 using the full-band Her X-1 spectral model. We note that the given uncertainties are purely statistical (provided at $1\sigma$ significance). \label{res_timeavg_full}}
\begin{center}
\begin{tabular}{ccccccc}
\hline
\hline
Component & Parameter & Unit & Comment & Orbit 1 & Orbit 2 & Orbit 3 \\
\hline 
Disk Wind & $\log N_{\rm{Fe~XXV}} $&\pcm & Fe XXV column density&$20.80_{-0.05}^{+0.04}$&$20.36_{-0.09}^{+0.08}$&$20.31_{-0.13}^{+0.10}$\\
\textsc{slab} & $\log N_{\rm{Fe~XXVI}} $ &\pcm & Fe XXVI column density&$21.85_{-0.02}^{+0.02}$&$21.61_{-0.02}^{+0.02}$&$21.47_{-0.02}^{+0.02}$\\
 &$v$& km/s & Velocity width&$281_{-11}^{+11}$&$326_{-16}^{+17}$&$370_{-30}^{+30}$\\
  &$z$& km/s & Outflow velocity&$336_{-13}^{+14}$&$550_{-20}^{+20}$&$690_{-40}^{+40}$\\
\hline
Primary continuum &$\xi$& &Photon escape time parameter  &$1.48_{-0.01}^{+0.01}$&$1.44_{-0.01}^{+0.01}$&$1.48_{-0.01}^{+0.01}$\\
\textsc{bwcycl} &$\delta$&& Bulk to thermal Compt. ratio&$1.666_{-0.004}^{+0.004}$&$1.860_{-0.005}^{+0.005}$&$1.803_{-0.004}^{+0.004}$\\
 &$T_{e}$& keV&Electron temperature &$5.749_{-0.005}^{+0.005}$&$5.645_{-0.005}^{+0.005}$&$5.702_{-0.005}^{+0.005}$\\
  &$r_0$& m & Column radius &$73.08_{-0.04}^{+0.04}$&$63.33_{-0.03}^{+0.03}$&$63.40_{-0.03}^{+0.03}$\\
\hline
Cyclotron Line&$E$&keV & Line energy &$35.74_{-0.08}^{+0.08}$&$35.71_{-0.08}^{+0.08}$&$36.07_{-0.08}^{+0.09}$\\
\textsc{gabs}&$\sigma$&keV & Line width &$6.93_{-0.04}^{+0.05}$&$6.88_{-0.04}^{+0.04}$&$7.21_{-0.04}^{+0.04}$\\
 &$d$&keV & Line depth parameter &$15.27_{-0.12}^{+0.12}$&$15.39_{-0.12}^{+0.12}$&$16.46_{-0.12}^{+0.12}$\\
\hline
Broad Fe K line&$E$& keV & Line energy&$6.38_{-0.03}^{+0.03}$&$6.52_{-0.02}^{+0.02}$&$6.50_{-0.02}^{+0.02}$\\
\textsc{gauss} &$\sigma$& keV &Line width &$1.09_{-0.02}^{+0.02}$&$1.05_{-0.02}^{+0.02}$&$1.07_{-0.02}^{+0.02}$\\
 &$\log(L)$&erg/s   &Line luminosity &$35.500_{-0.007}^{+0.007}$&$35.557_{-0.006}^{+0.006}$&$35.591_{-0.006}^{+0.006}$\\
\hline
Fe XXV line&$E$&keV & Line energy&$6.60_{-0.02}^{+0.02}$&$6.62_{-0.01}^{+0.01}$&$6.63_{-0.01}^{+0.01}$\\ 
\textsc{gauss} &$\sigma$& keV&Line width &$0.25_{-0.01}^{+0.01}$&$0.24_{-0.01}^{+0.01}$&$0.25_{-0.01}^{+0.01}$\\
 &$\log(L)$&erg/s&Line luminosity  &$34.82_{-0.02}^{+0.02}$&$34.93_{-0.01}^{+0.01}$&$35.00_{-0.01}^{+0.01}$\\
\hline
Fe I K$\alpha$ line &$E$&keV & Line energy&$6.400_{-0.001}^{+0.001}$&$6.401_{-0.001}^{+0.001}$&$6.402_{-0.001}^{+0.001}$\\
\textsc{gauss} &$\sigma$&keV &Line width &$0.014_{-0.001}^{+0.001}$&$0.015_{-0.001}^{+0.001}$&$0.016_{-0.001}^{+0.001}$\\
 &$\log(L)$&erg/s  & Line luminosity&$34.30_{-0.02}^{+0.02}$&$34.27_{-0.02}^{+0.02}$&$34.31_{-0.02}^{+0.02}$\\
\hline
Fe I K$\beta$ line&$E$&keV&Line energy & \multicolumn{3}{c}{Coupled with Fe I K$\alpha$ energy} \\
\textsc{gauss} &$\sigma$&keV&Line width &\multicolumn{3}{c}{Coupled with Fe I K$\alpha$ width}\\
&$\log(L)$&erg/s &Line luminosity &$33.44_{-0.13}^{+0.10}$&$33.48_{-0.10}^{+0.08}$&$33.17_{-0.34}^{+0.17}$\\
\hline
\hline
\end{tabular}
\end{center}
\end{table*}

\section{Full results of pulse-phase-resolved broad Fe K analysis of all orbits}
\label{app:orb1_orb3}

This appendix contains the full broad Fe K line pulse-phase-resolved modeling results for each orbit. Table \ref{res_Orb1} shows the results for Orbit 1, Table \ref{res_Orb2} the results for Orbit 2, and Table \ref{res_Orb3} the results for Orbit 3.

\begin{table*}
\caption{Results of the broad Fe K line modeling using the full-band Her X-1 spectral model from section \ref{sec:results} for Orbit 1. \label{res_Orb1}}
\begin{center}
\begin{tabular}{ccccc}
\hline
\hline
\multicolumn{5}{c}{Orbit 1}\\
\hline
\hline
Pulse Bin&Pulse Phase&Line Luminosity&Line Energy&Line $1\sigma$ width\\
&&$10^{35}$ erg/s&keV&keV\\
\hline
0 & 0.00  & $ 1.58^{+0.13}_{-0.13} $ & $ 6.33^{+0.08}_{-0.08} $ & $ 0.70^{+0.07}_{-0.06} $\\
1 & 0.05  & $ 2.14^{+0.12}_{-0.19} $ & $ 6.34^{+0.09}_{-0.13} $ & $ 1.06^{+0.12}_{-0.07} $\\
2 & 0.10  & $ 3.34^{+0.31}_{-0.29} $ & $ 6.16^{+0.07}_{-0.09} $ & $ 1.12^{+0.12}_{-0.08} $\\
3 & 0.14  & $ 2.73^{+0.14}_{-0.14} $ & $ 6.23^{+0.07}_{-0.07} $ & $ 1.01^{+0.06}_{-0.06} $\\
4 & 0.19  & $ 2.37^{+0.13}_{-0.13} $ & $ 6.26^{+0.08}_{-0.08} $ & $ 1.12^{+0.07}_{-0.07} $\\
5 & 0.24   & $ 2.46^{+0.11}_{-0.11} $ & $ 6.43^{+0.07}_{-0.07} $ & $ 1.15^{+0.06}_{-0.06} $\\
6 & 0.29   & $ 2.82^{+0.11}_{-0.11} $ & $ 6.33^{+0.07}_{-0.07} $ & $ 1.17^{+0.06}_{-0.05} $\\
7 & 0.33   & $ 3.33^{+0.12}_{-0.12} $ & $ 6.39^{+0.06}_{-0.06} $ & $ 1.14^{+0.05}_{-0.05} $\\
8 & 0.38   & $ 3.57^{+0.14}_{-0.14} $ & $ 6.35^{+0.06}_{-0.06} $ & $ 1.19^{+0.05}_{-0.05} $\\
9 & 0.43   & $ 3.44^{+0.12}_{-0.14} $ & $ 6.48^{+0.05}_{-0.05} $ & $ 1.09^{+0.05}_{-0.05} $\\
10 & 0.48  & $ 3.00^{+0.11}_{-0.11} $ & $ 6.43^{+0.06}_{-0.05} $ & $ 1.06^{+0.05}_{-0.05} $\\
11 & 0.52  & $ 2.79^{+0.10}_{-0.10} $ & $ 6.47^{+0.05}_{-0.05} $ & $ 0.91^{+0.04}_{-0.04} $\\
12 & 0.57  & $ 2.74^{+0.10}_{-0.10} $ & $ 6.40^{+0.05}_{-0.05} $ & $ 0.95^{+0.05}_{-0.04} $\\
13 & 0.62  & $ 2.72^{+0.11}_{-0.11} $ & $ 6.43^{+0.06}_{-0.06} $ & $ 1.02^{+0.05}_{-0.05} $\\
14 & 0.67  & $ 3.65^{+0.12}_{-0.13} $ & $ 6.36^{+0.06}_{-0.06} $ & $ 1.21^{+0.05}_{-0.05} $\\
15 & 0.71  & $ 4.78^{+0.14}_{-0.14} $ & $ 6.32^{+0.05}_{-0.05} $ & $ 1.28^{+0.05}_{-0.04} $\\
16 & 0.76  & $ 6.95^{+0.53}_{-0.42} $ & $ 6.33^{+0.05}_{-0.06} $ & $ 1.43^{+0.10}_{-0.07} $\\
17 & 0.81  & $ 6.94^{+0.69}_{-0.59} $ & $ 6.31^{+0.08}_{-0.08} $ & $ 1.39^{+0.11}_{-0.11} $\\
18 & 0.86  & $ 4.18^{+0.24}_{-0.24} $ & $ 6.49^{+0.06}_{-0.06} $ & $ 1.02^{+0.06}_{-0.03} $\\
19 & 0.90  & $ 2.30^{+0.16}_{-0.16} $ & $ 6.40^{+0.07}_{-0.07} $ & $ 0.76^{+0.06}_{-0.05} $\\
20 & 0.95  & $ 1.37^{+0.23}_{-0.27} $ & $ 6.39^{+0.11}_{-0.12} $ & $ 0.67^{+0.11}_{-0.08} $\\
\hline
\end{tabular}
\end{center}
\end{table*}

\begin{table*}
\caption{Results of the broad Fe K line modeling using the full-band Her X-1 spectral model from section \ref{sec:results} for Orbit 2. \label{res_Orb2}}
\begin{center}
\begin{tabular}{ccccc}
\hline
\hline
\multicolumn{5}{c}{Orbit 2}\\
\hline
\hline
Pulse Bin&Pulse Phase&Line Luminosity&Line Energy&Line $1\sigma$ width\\
&&$10^{35}$ erg/s&keV&keV\\
\hline
0 & 0.00      & $ 2.18^{+0.11}_{-0.12} $ & $ 6.52^{+0.05}_{-0.05} $ & $ 0.63^{+0.04}_{-0.04} $\\ 
1 & 0.05 & $ 2.60^{+0.12}_{-0.12} $ & $ 6.45^{+0.05}_{-0.05} $ & $ 0.76^{+0.04}_{-0.04} $\\
2 & 0.10 & $ 5.67^{+0.19}_{-0.19} $ & $ 6.19^{+0.06}_{-0.06} $ & $ 1.38^{+0.05}_{-0.05} $\\
3 & 0.14  & $ 6.01^{+0.61}_{-0.52} $ & $ 6.19^{+0.09}_{-0.10} $ & $ 1.51^{+0.12}_{-0.11} $\\
4 & 0.19  & $ 4.10^{+0.14}_{-0.15} $ & $ 6.44^{+0.06}_{-0.06} $ & $ 1.34^{+0.06}_{-0.05} $\\
5 & 0.24  & $ 3.03^{+0.10}_{-0.09} $ & $ 6.53^{+0.05}_{-0.06} $ & $ 1.20^{+0.05}_{-0.05} $\\
6 & 0.29  & $ 2.79^{+0.09}_{-0.09} $ & $ 6.55^{+0.05}_{-0.05} $ & $ 1.05^{+0.04}_{-0.04} $\\
7 & 0.33  & $ 3.13^{+0.10}_{-0.10} $ & $ 6.57^{+0.04}_{-0.04} $ & $ 0.95^{+0.04}_{-0.04} $\\
8 & 0.38   & $ 3.45^{+0.23}_{-0.20} $ & $ 6.58^{+0.04}_{-0.04} $ & $ 0.86^{+0.05}_{-0.08} $\\
9 & 0.43   & $ 3.43^{+0.11}_{-0.11} $ & $ 6.58^{+0.04}_{-0.04} $ & $ 1.00^{+0.04}_{-0.04} $\\
10 & 0.48  & $ 3.03^{+0.09}_{-0.12} $ & $ 6.54^{+0.04}_{-0.05} $ & $ 0.98^{+0.05}_{-0.04} $\\
11 & 0.52  & $ 3.06^{+0.10}_{-0.10} $ & $ 6.51^{+0.05}_{-0.04} $ & $ 1.01^{+0.04}_{-0.04} $\\
12 & 0.57  & $ 3.14^{+0.10}_{-0.10} $ & $ 6.49^{+0.05}_{-0.05} $ & $ 1.08^{+0.04}_{-0.04} $\\
13 & 0.62  & $ 3.48^{+0.19}_{-0.22} $ & $ 6.46^{+0.05}_{-0.05} $ & $ 1.16^{+0.08}_{-0.06} $\\
14 & 0.67  & $ 4.30^{+0.12}_{-0.11} $ & $ 6.44^{+0.04}_{-0.04} $ & $ 1.17^{+0.04}_{-0.04} $\\
15 & 0.71  & $ 5.03^{+0.23}_{-0.22} $ & $ 6.44^{+0.04}_{-0.04} $ & $ 1.14^{+0.07}_{-0.04} $\\
16 & 0.76  & $ 5.99^{+0.14}_{-0.15} $ & $ 6.44^{+0.04}_{-0.04} $ & $ 1.17^{+0.03}_{-0.03} $\\
17 & 0.81 & $ 6.54^{+0.16}_{-0.16} $ & $ 6.50^{+0.05}_{-0.05} $ & $ 1.36^{+0.05}_{-0.04} $\\
18 & 0.86  & $ 4.21^{+0.41}_{-0.38} $ & $ 6.69^{+0.09}_{-0.08} $ & $ 1.14^{+0.11}_{-0.10} $\\
19 & 0.90 & $ 2.50^{+0.15}_{-0.16} $ & $ 6.70^{+0.08}_{-0.08} $ & $ 0.92^{+0.06}_{-0.05} $\\
20 & 0.95  & $ 1.87^{+0.15}_{-0.17} $ & $ 6.60^{+0.05}_{-0.05} $ & $ 0.57^{+0.06}_{-0.05} $\\
\hline
\end{tabular}
\end{center}
\end{table*}

\begin{table*}
\caption{Results of the broad Fe K line modeling using the full-band Her X-1 spectral model from section \ref{sec:results} for Orbit 3. \label{res_Orb3}}
\begin{center}
\begin{tabular}{ccccc}
\hline
\hline
\multicolumn{5}{c}{Orbit 3}\\
\hline
\hline
Pulse Bin&Pulse Phase&Line Luminosity&Line Energy&Line $1\sigma$ width\\
&&$10^{35}$ erg/s&keV&keV\\
\hline
0 & 0.00   & $ 2.51^{+0.15}_{-0.15} $ & $ 6.40^{+0.08}_{-0.08} $ & $ 0.89^{+0.06}_{-0.06} $\\
1 & 0.05   & $ 8.5^{+1.5}_{-1.2} $ & $ 5.76^{+0.21}_{-0.23} $ & $ 1.79^{+0.22}_{-0.22} $\\
2 & 0.10   & $ 6.51^{+0.28}_{-0.36} $ & $ 5.88^{+0.08}_{-0.10} $ & $ 1.56^{+0.08}_{-0.07} $\\
3 & 0.14   & $ 4.37^{+0.44}_{-0.38} $ & $ 6.17^{+0.09}_{-0.09} $ & $ 1.29^{+0.10}_{-0.10} $\\
4 & 0.19   & $ 3.08^{+0.39}_{-0.33} $ & $ 6.42^{+0.10}_{-0.09} $ & $ 1.29^{+0.12}_{-0.13} $\\
5 & 0.24  & $ 2.94^{+0.12}_{-0.13} $ & $ 6.66^{+0.07}_{-0.07} $ & $ 1.26^{+0.06}_{-0.07} $\\
6 & 0.29   & $ 3.65^{+0.14}_{-0.14} $ & $ 6.58^{+0.06}_{-0.06} $ & $ 1.24^{+0.05}_{-0.05} $\\
7 & 0.33  & $ 4.41^{+0.14}_{-0.13} $ & $ 6.48^{+0.05}_{-0.05} $ & $ 1.11^{+0.04}_{-0.04} $ \\
8 & 0.38  & $ 4.53^{+0.33}_{-0.33} $ & $ 6.48^{+0.05}_{-0.05} $ & $ 1.01^{+0.07}_{-0.07} $\\
9 & 0.43  & $ 4.10^{+0.12}_{-0.15} $ & $ 6.53^{+0.04}_{-0.04} $ & $ 0.96^{+0.04}_{-0.03} $\\
10 & 0.48 & $ 3.64^{+0.14}_{-0.14} $ & $ 6.55^{+0.05}_{-0.05} $ & $ 1.11^{+0.05}_{-0.05} $\\
11 & 0.52 & $ 3.91^{+0.13}_{-0.13} $ & $ 6.50^{+0.06}_{-0.06} $ & $ 1.18^{+0.05}_{-0.05} $\\
12 & 0.57 & $ 3.99^{+0.13}_{-0.13} $ & $ 6.44^{+0.05}_{-0.05} $ & $ 1.09^{+0.04}_{-0.04} $\\
13 & 0.62 & $ 4.26^{+0.13}_{-0.13} $ & $ 6.45^{+0.04}_{-0.04} $ & $ 1.04^{+0.04}_{-0.04} $\\
14 & 0.67 & $ 5.04^{+0.14}_{-0.14} $ & $ 6.51^{+0.04}_{-0.04} $ & $ 1.17^{+0.04}_{-0.04} $\\
15 & 0.71 & $ 5.72^{+0.16}_{-0.18} $ & $ 6.53^{+0.05}_{-0.05} $ & $ 1.33^{+0.05}_{-0.05} $\\
16 & 0.76 & $ 5.87^{+0.19}_{-0.19} $ & $ 6.59^{+0.06}_{-0.06} $ & $ 1.35^{+0.05}_{-0.05} $\\
17 & 0.81  & $ 4.30^{+0.18}_{-0.18} $ & $ 6.69^{+0.07}_{-0.07} $ & $ 1.17^{+0.06}_{-0.06} $\\
18 & 0.86  & $ 3.17^{+0.15}_{-0.17} $ & $ 6.63^{+0.06}_{-0.06} $ & $ 0.81^{+0.05}_{-0.04} $\\
19 & 0.90  & $ 2.33^{+0.15}_{-0.15} $ & $ 6.58^{+0.06}_{-0.06} $ & $ 0.66^{+0.05}_{-0.04} $\\
20 & 0.95  & $ 2.41^{+0.15}_{-0.16} $ & $ 6.50^{+0.06}_{-0.06} $ & $ 0.75^{+0.05}_{-0.05} $\\
\hline
\end{tabular}
\end{center}
\end{table*}

\section{Comparison of full-band complex model analysis with a narrow-band simplified analysis}
\label{app:narrowband}

\begin{figure*}
\begin{center}
\includegraphics[width=0.55\textwidth]{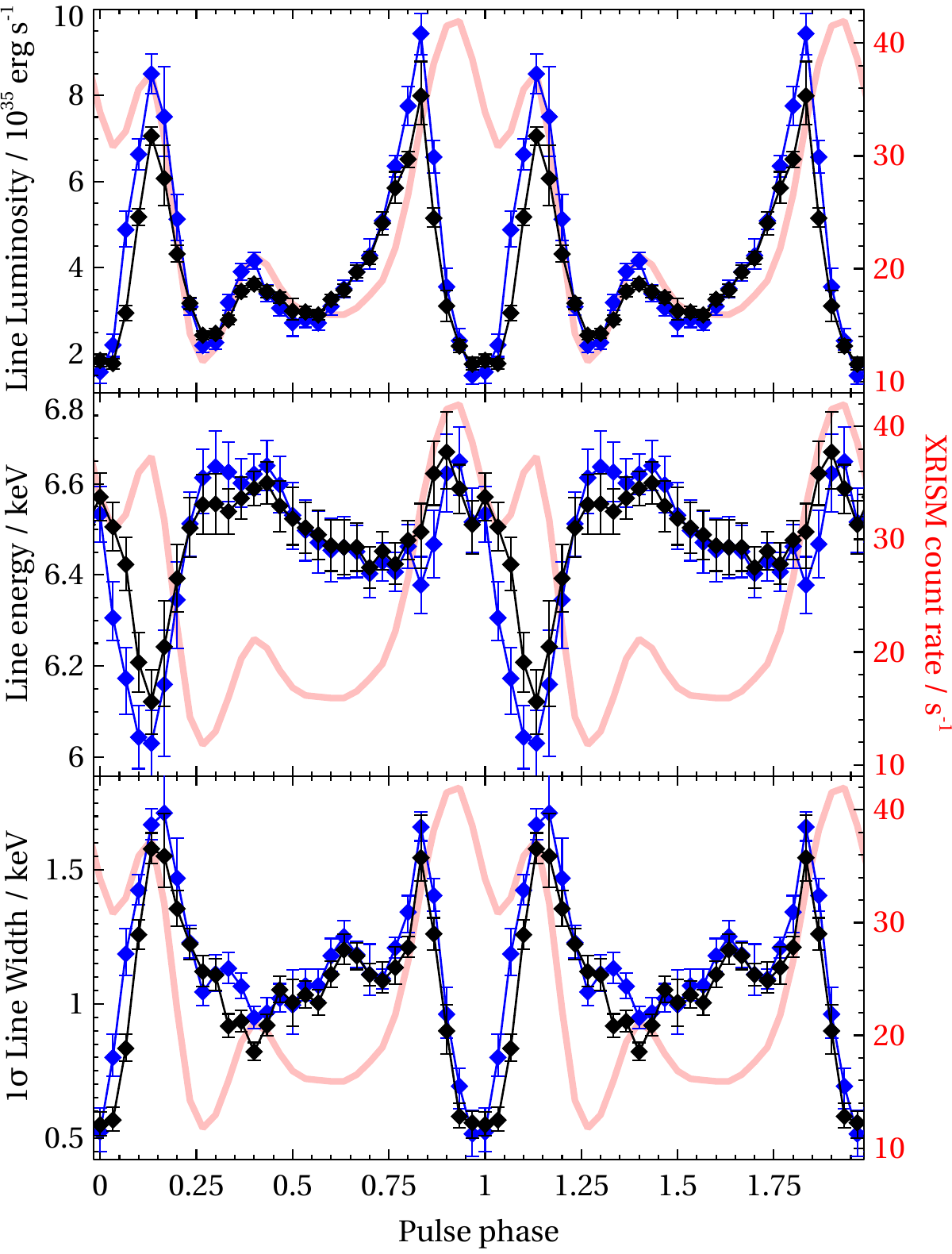}
\caption{Comparison of the best-fitting broad Fe K line properties with pulse phase during Orbit 2 using the full spectral model covering $2-75$ keV (black, identical to Fig. \ref{Fig_orb2_results}) and a simplified phenomenological continuum model applied to a narrower energy band of $2-12$ keV (blue). The Her X-1 continuum pulsation is also shown for comparison using the \xrism\ count rate versus pulse phase (red).  \label{Full_vs_simple_analysis}}
\end{center}
\end{figure*}

We compare our full-band, physically-motivated continuum model with a simpler, narrower energy band model, to assess the effect of model and dataset choice on the best-fitting broad Fe K emission line properties. For this narrow-band spectral fit, we only use Hp quality \xrism/Resolve data in the $2-12$ keV emission band (such that the Fe K band is roughly the center of the analyzed emission range), and do not use hard X-ray \nustar\ data. Additionally, we simplified the continuum emission model. Instead of the \textsc{bwcycl} full-band model, we use a simple power law, and do not require the inclusion of a cyclotron resonance scattering feature which only affects the X-ray spectrum above 12 keV. The Fe K emission line modeling, ionized disk wind modeling and the neutral absorption remains the same as in the full model. The final model is in symbolic form therefore: \textsc{tbabs $\times$ slab $\times$ (po + 4 $\times$ gauss)}. The comparison between the full-band and simplified analysis for the highest S/N Orbit 2 is shown in Fig. \ref{Full_vs_simple_analysis}. 

We note that for this comparison we used a finer time sampling of 30 bins, each with a width of 0.1 times the pulse period, which is possible for Orbit 2 thanks to its high S/N. We find that the evolution of all broad Fe K line parameters is qualitatively comparable in both analyses. We observed the same two, roughly equal peaks in both the line luminosity and line width pulse evolution. The same sharp drop in the line centroid is observed at roughly the same pulse phase in both approaches. However, some differences are present as well. The line luminosity peaks at higher values for the simplified analysis, and the centroid energy evolution appears to be shifted by a small amount in pulse phase for certain pulse bins around the sharp drop in energy. Nevertheless, the results are broadly consistent and we conclude that to first order, our analysis is not strongly biased by the usage of the complex full-band spectral model.


\bibliography{References}{}

@INPROCEEDINGS{Kaastra+96,
       author = {{Kaastra}, J.~S. and {Mewe}, R. and {Nieuwenhuijzen}, H.},
        title = "{SPEX: a new code for spectral analysis of X \& UV spectra.}",
    booktitle = {UV and X-ray Spectroscopy of Astrophysical and Laboratory Plasmas},
         year = 1996,
        month = jan,
        pages = {411-414},
       adsurl = {https://ui.adsabs.harvard.edu/abs/1996uxsa.conf..411K}
}

@INPROCEEDINGS{Arnaud+96,
       author = {{Arnaud}, K.~A.},
        title = "{XSPEC: The First Ten Years}",
    booktitle = {Astronomical Data Analysis Software and Systems V},
         year = 1996,
       editor = {{Jacoby}, George H. and {Barnes}, Jeannette},
       series = {Astronomical Society of the Pacific Conference Series},
       volume = {101},
        month = jan,
        pages = {17},
       adsurl = {https://ui.adsabs.harvard.edu/abs/1996ASPC..101...17A}
}

@ARTICLE{Cash+79,
       author = {{Cash}, W.},
        title = "{Parameter estimation in astronomy through application of the likelihood ratio.}",
      journal = {\apj},
         year = 1979,
        month = mar,
       volume = {228},
        pages = {939-947},
          doi = {10.1086/156922},
       adsurl = {https://ui.adsabs.harvard.edu/abs/1979ApJ...228..939C}
}

@ARTICLE{Kosec+20,
       author = {{Kosec}, P. and {Fabian}, A.~C. and {Pinto}, C. and {Walton}, D.~J. and {Dyda}, S. and {Reynolds}, C.~S.},
        title = "{An ionized accretion disc wind in Hercules X-1}",
      journal = {\mnras},
         year = 2020,
        month = jan,
       volume = {491},
       number = {3},
        pages = {3730-3750},
          doi = {10.1093/mnras/stz3200},
archivePrefix = {arXiv},
       eprint = {1910.08337},
 primaryClass = {astro-ph.HE},
       adsurl = {https://ui.adsabs.harvard.edu/abs/2020MNRAS.491.3730K}
}

@ARTICLE{Leahy+14,
       author = {{Leahy}, D.~A. and {Abdallah}, M.~H.},
        title = "{HZ Her: Stellar Radius from X-Ray Eclipse Observations, Evolutionary State, and a New Distance}",
      journal = {\apj},
         year = 2014,
        month = oct,
       volume = {793},
       number = {2},
          eid = {79},
        pages = {79},
          doi = {10.1088/0004-637X/793/2/79},
archivePrefix = {arXiv},
       eprint = {1406.6138},
 primaryClass = {astro-ph.SR},
       adsurl = {https://ui.adsabs.harvard.edu/abs/2014ApJ...793...79L}
}

@ARTICLE{Tananbaum+72,
       author = {{Tananbaum}, H. and {Gursky}, H. and {Kellogg}, E.~M. and {Levinson}, R. and {Schreier}, E. and {Giacconi}, R.},
        title = "{Discovery of a Periodic Pulsating Binary X-Ray Source in Hercules from UHURU}",
      journal = {\apjl},
         year = 1972,
        month = jun,
       volume = {174},
        pages = {L143},
          doi = {10.1086/180968},
       adsurl = {https://ui.adsabs.harvard.edu/abs/1972ApJ...174L.143T}
}

@ARTICLE{Gerend+76,
       author = {{Gerend}, D. and {Boynton}, P.~E.},
        title = "{Optical clues to the nature of Hercules X-1 / HZ Herculis.}",
      journal = {\apj},
         year = 1976,
        month = oct,
       volume = {209},
        pages = {562-573},
          doi = {10.1086/154751},
       adsurl = {https://ui.adsabs.harvard.edu/abs/1976ApJ...209..562G}
}

@ARTICLE{Katz+73,
       author = {{Katz}, J.~I.},
        title = "{Thirty-five-day Periodicity in Her X-1}",
      journal = {Nature Physical Science},
         year = 1973,
        month = dec,
       volume = {246},
       number = {154},
        pages = {87-89},
          doi = {10.1038/physci246087a0},
       adsurl = {https://ui.adsabs.harvard.edu/abs/1973NPhS..246...87K}
}

@ARTICLE{Kosec+23a,
       author = {{Kosec}, P. and {Kara}, E. and {Fabian}, A. and {F{\"u}rst}, F. and {Pinto}, C. and {Psaradaki}, I. and {Reynolds}, C.~S. and {Rogantini}, D. and {Walton}, D.~J. and {Ballhausen}, R. and {Canizares}, C. and {Dyda}, S. and {Staubert}, R. and {Wilms}, J.},
        title = "{Vertical wind structure in an X-ray binary revealed by a precessing accretion disk}",
      journal = {Nature Astronomy},
         year = 2023,
        month = jun,
       volume = {7},
        pages = {715-723},
          doi = {10.1038/s41550-023-01929-7},
archivePrefix = {arXiv},
       eprint = {2304.05490},
 primaryClass = {astro-ph.HE},
       adsurl = {https://ui.adsabs.harvard.edu/abs/2023NatAs...7..715K}
}

@ARTICLE{Tashiro+22,
       author = {{Tashiro}, Makoto S.},
        title = "{XRISM: X-ray imaging and spectroscopy mission}",
      journal = {International Journal of Modern Physics D},
         year = 2022,
        month = jan,
       volume = {31},
       number = {2},
          eid = {2230001},
        pages = {2230001},
          doi = {10.1142/S0218271822300014},
       adsurl = {https://ui.adsabs.harvard.edu/abs/2022IJMPD..3130001T}
}

@ARTICLE{Kosec+24,
       author = {{Kosec}, P. and {Rogantini}, D. and {Kara}, E. and {Canizares}, C.~R. and {Fabian}, A.~C. and {Pinto}, C. and {Psaradaki}, I. and {Staubert}, R. and {Walton}, D.~J.},
        title = "{Constraining the Number Density of the Accretion Disk Wind in Hercules X-1 Using Its Ionization Response to X-Ray Pulsations}",
      journal = {\apj},
         year = 2024,
        month = sep,
       volume = {972},
       number = {1},
          eid = {32},
        pages = {32},
          doi = {10.3847/1538-4357/ad5b5a},
archivePrefix = {arXiv},
       eprint = {2401.00754},
 primaryClass = {astro-ph.HE},
       adsurl = {https://ui.adsabs.harvard.edu/abs/2024ApJ...972...32K}
}

@ARTICLE{Kaastra+16,
       author = {{Kaastra}, J.~S. and {Bleeker}, J.~A.~M.},
        title = "{Optimal binning of X-ray spectra and response matrix design}",
      journal = {\aap},
         year = 2016,
        month = mar,
       volume = {587},
          eid = {A151},
        pages = {A151},
          doi = {10.1051/0004-6361/201527395},
archivePrefix = {arXiv},
       eprint = {1601.05309},
 primaryClass = {astro-ph.IM},
       adsurl = {https://ui.adsabs.harvard.edu/abs/2016A&A...587A.151K}
}

@ARTICLE{Harrison+13,
       author = {{Harrison}, Fiona A. and {Craig}, William W. and {Christensen}, Finn E. and {Hailey}, Charles J. and {Zhang}, William W. and {Boggs}, Steven E. and {Stern}, Daniel and {Cook}, W. Rick and {Forster}, Karl and {Giommi}, Paolo and {Grefenstette}, Brian W. and {Kim}, Yunjin and {Kitaguchi}, Takao and {Koglin}, Jason E. and {Madsen}, Kristin K. and {Mao}, Peter H. and {Miyasaka}, Hiromasa and {Mori}, Kaya and {Perri}, Matteo and {Pivovaroff}, Michael J. and {Puccetti}, Simonetta and {Rana}, Vikram R. and {Westergaard}, Niels J. and {Willis}, Jason and {Zoglauer}, Andreas and {An}, Hongjun and {Bachetti}, Matteo and {Barri{\`e}re}, Nicolas M. and {Bellm}, Eric C. and {Bhalerao}, Varun and {Brejnholt}, Nicolai F. and {Fuerst}, Felix and {Liebe}, Carl C. and {Markwardt}, Craig B. and {Nynka}, Melania and {Vogel}, Julia K. and {Walton}, Dominic J. and {Wik}, Daniel R. and {Alexander}, David M. and {Cominsky}, Lynn R. and {Hornschemeier}, Ann E. and {Hornstrup}, Allan and {Kaspi}, Victoria M. and {Madejski}, Greg M. and {Matt}, Giorgio and {Molendi}, Silvano and {Smith}, David M. and {Tomsick}, John A. and {Ajello}, Marco and {Ballantyne}, David R. and {Balokovi{\'c}}, Mislav and {Barret}, Didier and {Bauer}, Franz E. and {Blandford}, Roger D. and {Brandt}, W. Niel and {Brenneman}, Laura W. and {Chiang}, James and {Chakrabarty}, Deepto and {Chenevez}, Jerome and {Comastri}, Andrea and {Dufour}, Francois and {Elvis}, Martin and {Fabian}, Andrew C. and {Farrah}, Duncan and {Fryer}, Chris L. and {Gotthelf}, Eric V. and {Grindlay}, Jonathan E. and {Helfand}, David J. and {Krivonos}, Roman and {Meier}, David L. and {Miller}, Jon M. and {Natalucci}, Lorenzo and {Ogle}, Patrick and {Ofek}, Eran O. and {Ptak}, Andrew and {Reynolds}, Stephen P. and {Rigby}, Jane R. and {Tagliaferri}, Gianpiero and {Thorsett}, Stephen E. and {Treister}, Ezequiel and {Urry}, C. Megan},
        title = "{The Nuclear Spectroscopic Telescope Array (NuSTAR) High-energy X-Ray Mission}",
      journal = {\apj},
         year = 2013,
        month = jun,
       volume = {770},
       number = {2},
          eid = {103},
        pages = {103},
          doi = {10.1088/0004-637X/770/2/103},
archivePrefix = {arXiv},
       eprint = {1301.7307},
 primaryClass = {astro-ph.IM},
       adsurl = {https://ui.adsabs.harvard.edu/abs/2013ApJ...770..103H}
}

@ARTICLE{Jansen+01,
       author = {{Jansen}, F. and {Lumb}, D. and {Altieri}, B. and {Clavel}, J. and {Ehle}, M. and {Erd}, C. and {Gabriel}, C. and {Guainazzi}, M. and {Gondoin}, P. and {Much}, R. and {Munoz}, R. and {Santos}, M. and {Schartel}, N. and {Texier}, D. and {Vacanti}, G.},
        title = "{XMM-Newton observatory. I. The spacecraft and operations}",
      journal = {\aap},
         year = 2001,
        month = jan,
       volume = {365},
        pages = {L1-L6},
          doi = {10.1051/0004-6361:20000036},
       adsurl = {https://ui.adsabs.harvard.edu/abs/2001A&A...365L...1J}
}

@ARTICLE{Kosec+22,
       author = {{Kosec}, P. and {Kara}, E. and {Fabian}, A.~C. and {F{\"u}rst}, F. and {Pinto}, C. and {Psaradaki}, I. and {Reynolds}, C.~S. and {Rogantini}, D. and {Walton}, D.~J. and {Ballhausen}, R. and {Canizares}, C. and {Dyda}, S. and {Staubert}, R. and {Wilms}, J.},
        title = "{The Long Stare at Hercules X-1. I. Emission Lines from the Outer Disk, the Magnetosphere Boundary, and the Accretion Curtain}",
      journal = {\apj},
         year = 2022,
        month = sep,
       volume = {936},
       number = {2},
          eid = {185},
        pages = {185},
          doi = {10.3847/1538-4357/ac897e},
archivePrefix = {arXiv},
       eprint = {2208.08930},
 primaryClass = {astro-ph.HE},
       adsurl = {https://ui.adsabs.harvard.edu/abs/2022ApJ...936..185K}
}

@ARTICLE{Truemper+78,
       author = {{Truemper}, J. and {Pietsch}, W. and {Reppin}, C. and {Voges}, W. and {Staubert}, R. and {Kendziorra}, E.},
        title = "{Evidence for strong cyclotron line emission in the hard X-ray spectrum of Hercules X-1.}",
      journal = {\apjl},
         year = 1978,
        month = feb,
       volume = {219},
        pages = {L105-L110},
          doi = {10.1086/182617},
       adsurl = {https://ui.adsabs.harvard.edu/abs/1978ApJ...219L.105T}
}

@ARTICLE{Kaastra+02,
       author = {{Kaastra}, J.~S. and {Steenbrugge}, K.~C. and {Raassen}, A.~J.~J. and {van der Meer}, R.~L.~J. and {Brinkman}, A.~C. and {Liedahl}, D.~A. and {Behar}, E. and {de Rosa}, A.},
        title = "{X-ray spectroscopy of NGC 5548}",
      journal = {\aap},
         year = 2002,
        month = may,
       volume = {386},
        pages = {427-445},
          doi = {10.1051/0004-6361:20020235},
archivePrefix = {arXiv},
       eprint = {astro-ph/0202481},
 primaryClass = {astro-ph},
       adsurl = {https://ui.adsabs.harvard.edu/abs/2002A&A...386..427K}
}

@ARTICLE{Jimenez+05,
       author = {{Jimenez-Garate}, M.~A. and {Raymond}, J.~C. and {Liedahl}, D.~A. and {Hailey}, C.~J.},
        title = "{Identification of an Extended Accretion Disk Corona in the Hercules X-1 Low State: Moderate Optical Depth, Precise Density Determination, and Verification of CNO Abundances}",
      journal = {\apj},
         year = 2005,
        month = jun,
       volume = {625},
       number = {2},
        pages = {931-950},
          doi = {10.1086/426702},
archivePrefix = {arXiv},
       eprint = {astro-ph/0411780},
 primaryClass = {astro-ph},
       adsurl = {https://ui.adsabs.harvard.edu/abs/2005ApJ...625..931J}
}

@ARTICLE{Jimenez+02,
       author = {{Jimenez-Garate}, M.~A. and {Hailey}, C.~J. and {den Herder}, J.~W. and {Zane}, S. and {Ramsay}, G.},
        title = "{High-Resolution X-Ray Spectroscopy of Hercules X-1 with the XMM-Newton Reflection Grating Spectrometer: CNO Element Abundance Measurements and Density Diagnostics of a Photoionized Plasma}",
      journal = {\apj},
         year = 2002,
        month = oct,
       volume = {578},
       number = {1},
        pages = {391-404},
          doi = {10.1086/342348},
archivePrefix = {arXiv},
       eprint = {astro-ph/0206181},
 primaryClass = {astro-ph},
       adsurl = {https://ui.adsabs.harvard.edu/abs/2002ApJ...578..391J}
}

@INPROCEEDINGS{Weisskopf+00,
       author = {{Weisskopf}, Martin C. and {Tananbaum}, Harvey D. and {Van Speybroeck}, Leon P. and {O'Dell}, Stephen L.},
        title = "{Chandra X-ray Observatory (CXO): overview}",
    booktitle = {X-Ray Optics, Instruments, and Missions III},
         year = 2000,
       editor = {{Truemper}, Joachim E. and {Aschenbach}, Bernd},
       series = {Society of Photo-Optical Instrumentation Engineers (SPIE) Conference Series},
       volume = {4012},
        month = jul,
        pages = {2-16},
          doi = {10.1117/12.391545},
archivePrefix = {arXiv},
       eprint = {astro-ph/0004127},
 primaryClass = {astro-ph},
       adsurl = {https://ui.adsabs.harvard.edu/abs/2000SPIE.4012....2W}
}

@ARTICLE{Scott+00,
       author = {{Scott}, D. Matthew and {Leahy}, Denis A. and {Wilson}, Robert B.},
        title = "{The 35 Day Evolution of the Hercules X-1 Pulse Profile: Evidence for a Resolved Inner Disk Occultation of the Neutron Star}",
      journal = {\apj},
         year = 2000,
        month = aug,
       volume = {539},
       number = {1},
        pages = {392-412},
          doi = {10.1086/309203},
archivePrefix = {arXiv},
       eprint = {astro-ph/0002327},
 primaryClass = {astro-ph},
       adsurl = {https://ui.adsabs.harvard.edu/abs/2000ApJ...539..392S}
}

@ARTICLE{Kosec+26,
       author = {{Kosec}, Peter and {Brenneman}, Laura and {Kara}, Erin and {Enoto}, Teruaki and {Narita}, Takuto and {Sakamoto}, Koh and {Staubert}, Rudiger and {Barra}, Francesco and {Fabian}, Andrew and {Miller}, Jon M. and {Pinto}, Ciro and {Rogantini}, Daniele and {Walton}, Dominic and {Nagai}, Yutaro},
        title = "{XRISM/Resolve observations of Hercules X-1: vertical structure and kinematics of the disk wind}",
      journal = {arXiv e-prints},
         year = 2025,
        month = oct,
          eid = {arXiv:2510.07615},
        pages = {arXiv:2510.07615},
          doi = {10.48550/arXiv.2510.07615},
archivePrefix = {arXiv},
       eprint = {2510.07615},
 primaryClass = {astro-ph.HE},
       adsurl = {https://ui.adsabs.harvard.edu/abs/2025arXiv251007615K}
}

@ARTICLE{Ramsay+02,
       author = {{Ramsay}, Gavin and {Zane}, Silvia and {Jimenez-Garate}, Mario A. and {den Herder}, Jan-Willem and {Hailey}, C.~J.},
        title = "{XMM-Newton EPIC observations of Her X-1}",
      journal = {\mnras},
         year = 2002,
        month = dec,
       volume = {337},
       number = {4},
        pages = {1185-1192},
          doi = {10.1046/j.1365-8711.2002.06002.x},
archivePrefix = {arXiv},
       eprint = {astro-ph/0209020},
 primaryClass = {astro-ph},
       adsurl = {https://ui.adsabs.harvard.edu/abs/2002MNRAS.337.1185R}
}

@ARTICLE{Vasco+13,
       author = {{Vasco}, D. and {Staubert}, R. and {Klochkov}, D. and {Santangelo}, A. and {Shakura}, N. and {Postnov}, K.},
        title = "{Pulse phase and precession phase resolved spectroscopy of Hercules X-1: studying a representative Main-On with RXTE}",
      journal = {\aap},
         year = 2013,
        month = feb,
       volume = {550},
          eid = {A111},
        pages = {A111},
          doi = {10.1051/0004-6361/201220181},
archivePrefix = {arXiv},
       eprint = {1301.3378},
 primaryClass = {astro-ph.HE},
       adsurl = {https://ui.adsabs.harvard.edu/abs/2013A&A...550A.111V}
}

@ARTICLE{Bykov+21,
       author = {{Bykov}, S.~D. and {Filippova}, E.~V. and {Gilfanov}, M.~R. and {Tsygankov}, S.~S. and {Lutovinov}, A.~A. and {Molkov}, S.~V.},
        title = "{Pulsating iron spectral features in the emission of X-ray pulsar V 0332+53}",
      journal = {\mnras},
         year = 2021,
        month = sep,
       volume = {506},
       number = {2},
        pages = {2156-2169},
          doi = {10.1093/mnras/stab1852},
archivePrefix = {arXiv},
       eprint = {2106.14261},
 primaryClass = {astro-ph.HE},
       adsurl = {https://ui.adsabs.harvard.edu/abs/2021MNRAS.506.2156B}
}

@ARTICLE{Roy+25,
       author = {{Roy}, Kinjal and {Manikantan}, Hemanth and {Paul}, Biswajit},
        title = "{XMM-Newton View of Pulsating Iron Fluorescent Emission from Centaurus X-3}",
      journal = {\apj},
         year = 2025,
        month = oct,
       volume = {991},
       number = {2},
          eid = {162},
        pages = {162},
          doi = {10.3847/1538-4357/adf8e6},
archivePrefix = {arXiv},
       eprint = {2509.25036},
 primaryClass = {astro-ph.HE},
       adsurl = {https://ui.adsabs.harvard.edu/abs/2025ApJ...991..162R}
}

@ARTICLE{Xiao+24,
       author = {{Xiao}, Y.~X. and {Xu}, Y.~J. and {Ge}, M.~Y. and {Lu}, F.~J. and {Zhang}, S.~N. and {Zhang}, S. and {Tao}, L. and {Qu}, J.~L. and {Wang}, P.~J. and {Kong}, L.~D. and {Tuo}, Y.~L. and {You}, Y. and {Zhao}, S.~J. and {Peng}, J.~Q. and {Du}, Y.~F. and {Zhang}, Y.~H. and {Ye}, W.~T.},
        title = "{Pulsed Iron Line Emission from the First Galactic Ultraluminous X-Ray Pulsar Swift J0243.6+6124}",
      journal = {\apj},
         year = 2024,
        month = apr,
       volume = {965},
       number = {1},
          eid = {18},
        pages = {18},
          doi = {10.3847/1538-4357/ad24f8},
archivePrefix = {arXiv},
       eprint = {2401.15992},
 primaryClass = {astro-ph.HE},
       adsurl = {https://ui.adsabs.harvard.edu/abs/2024ApJ...965...18X}
}

@ARTICLE{Staubert+19,
       author = {{Staubert}, R. and {Tr{\"u}mper}, J. and {Kendziorra}, E. and {Klochkov}, D. and {Postnov}, K. and {Kretschmar}, P. and {Pottschmidt}, K. and {Haberl}, F. and {Rothschild}, R.~E. and {Santangelo}, A. and {Wilms}, J. and {Kreykenbohm}, I. and {F{\"u}rst}, F.},
        title = "{Cyclotron lines in highly magnetized neutron stars}",
      journal = {\aap},
         year = 2019,
        month = feb,
       volume = {622},
          eid = {A61},
        pages = {A61},
          doi = {10.1051/0004-6361/201834479},
archivePrefix = {arXiv},
       eprint = {1812.03461},
 primaryClass = {astro-ph.HE},
       adsurl = {https://ui.adsabs.harvard.edu/abs/2019A&A...622A..61S}
}

@ARTICLE{Doroshenko+22,
       author = {{Doroshenko}, Victor and {Poutanen}, Juri and {Tsygankov}, Sergey S. and {Suleimanov}, Valery F. and {Bachetti}, Matteo and {Caiazzo}, Ilaria and {Costa}, Enrico and {Di Marco}, Alessandro and {Heyl}, Jeremy and {La Monaca}, Fabio and {Muleri}, Fabio and {Mushtukov}, Alexander A. and {Pavlov}, George G. and {Ramsey}, Brian D. and {Rankin}, John and {Santangelo}, Andrea and {Soffitta}, Paolo and {Staubert}, R{\"u}diger and {Weisskopf}, Martin C. and {Zane}, Silvia and {Agudo}, Iv{\'a}n and {Antonelli}, Lucio A. and {Baldini}, Luca and {Baumgartner}, Wayne H. and {Bellazzini}, Ronaldo and {Bianchi}, Stefano and {Bongiorno}, Stephen D. and {Bonino}, Raffaella and {Brez}, Alessandro and {Bucciantini}, Niccol{\`o} and {Capitanio}, Fiamma and {Castellano}, Simone and {Cavazzuti}, Elisabetta and {Ciprini}, Stefano and {De Rosa}, Alessandra and {Del Monte}, Ettore and {Di Gesu}, Laura and {Di Lalla}, Niccol{\`o} and {Donnarumma}, Immacolata and {Dov{\v{c}}iak}, Michal and {Ehlert}, Steven R. and {Enoto}, Teruaki and {Evangelista}, Yuri and {Fabiani}, Sergio and {Ferrazzoli}, Riccardo and {Garcia}, Javier A. and {Gunji}, Shuichi and {Hayashida}, Kiyoshi and {Iwakiri}, Wataru and {Jorstad}, Svetlana G. and {Karas}, Vladimir and {Kitaguchi}, Takao and {Kolodziejczak}, Jeffery J. and {Krawczynski}, Henric and {Latronico}, Luca and {Liodakis}, Ioannis and {Maldera}, Simone and {Manfreda}, Alberto and {Marin}, Fr{\'e}d{\'e}ric and {Marinucci}, Andrea and {Marscher}, Alan P. and {Marshall}, Herman L. and {Matt}, Giorgio and {Mitsuishi}, Ikuyuki and {Mizuno}, Tsunefumi and {Ng}, Chi-Yung and {O'Dell}, Stephen L. and {Omodei}, Nicola and {Oppedisano}, Chiara and {Papitto}, Alessandro and {Peirson}, Abel L. and {Perri}, Matteo and {Pesce-Rollins}, Melissa and {Pilia}, Maura and {Possenti}, Andrea and {Puccetti}, Simonetta and {Ratheesh}, Ajay and {Romani}, Roger W. and {Sgr{\`o}}, Carmelo and {Slane}, Patrick and {Spandre}, Gloria and {Sunyaev}, Rashid A. and {Tamagawa}, Toru and {Tavecchio}, Fabrizio and {Taverna}, Roberto and {Tawara}, Yuzuru and {Tennant}, Allyn F. and {Thomas}, Nicolas E. and {Tombesi}, Francesco and {Trois}, Alessio and {Turolla}, Roberto and {Vink}, Jacco and {Wu}, Kinwah and {Xie}, Fei},
        title = "{Determination of X-ray pulsar geometry with IXPE polarimetry}",
      journal = {Nature Astronomy},
         year = 2022,
        month = dec,
       volume = {6},
        pages = {1433-1443},
          doi = {10.1038/s41550-022-01799-5},
archivePrefix = {arXiv},
       eprint = {2206.07138},
 primaryClass = {astro-ph.HE},
       adsurl = {https://ui.adsabs.harvard.edu/abs/2022NatAs...6.1433D}
}

@ARTICLE{Heyl+24,
       author = {{Heyl}, Jeremy and {Doroshenko}, Victor and {Gonz{\'a}lez-Caniulef}, Denis and {Caiazzo}, Ilaria and {Poutanen}, Juri and {Mushtukov}, Alexander and {Tsygankov}, Sergey S. and {Kirmizibayrak}, Demet and {Bachetti}, Matteo and {Pavlov}, George G. and {Forsblom}, Sofia V. and {Malacaria}, Christian and {Suleimanov}, Valery F. and {Agudo}, Iv{\'a}n and {Antonelli}, Lucio Angelo and {Baldini}, Luca and {Baumgartner}, Wayne H. and {Bellazzini}, Ronaldo and {Bianchi}, Stefano and {Bongiorno}, Stephen D. and {Bonino}, Raffaella and {Brez}, Alessandro and {Bucciantini}, Niccol{\`o} and {Capitanio}, Fiamma and {Castellano}, Simone and {Cavazzuti}, Elisabetta and {Chen}, Chien-Ting and {Ciprini}, Stefano and {Costa}, Enrico and {De Rosa}, Alessandra and {Del Monte}, Ettore and {Di Gesu}, Laura and {Di Lalla}, Niccol{\`o} and {Di Marco}, Alessandro and {Donnarumma}, Immacolata and {Dov{\v{c}}iak}, Michal and {Ehlert}, Steven R. and {Enoto}, Teruaki and {Evangelista}, Yuri and {Fabiani}, Sergio and {Ferrazzoli}, Riccardo and {Garcia}, Javier A. and {Gunji}, Shuichi and {Hayashida}, Kiyoshi and {Iwakiri}, Wataru and {Jorstad}, Svetlana G. and {Kaaret}, Philip and {Karas}, Vladimir and {Kislat}, Fabian and {Kitaguchi}, Takao and {Kolodziejczak}, Jeffery J. and {Krawczynski}, Henric and {La Monaca}, Fabio and {Latronico}, Luca and {Liodakis}, Ioannis and {Maldera}, Simone and {Manfreda}, Alberto and {Marin}, Fr{\'e}d{\'e}ric and {Marinucci}, Andrea and {Marscher}, Alan P. and {Marshall}, Herman L. and {Massaro}, Francesco and {Matt}, Giorgio and {Mitsuishi}, Ikuyuki and {Mizuno}, Tsunefumi and {Muleri}, Fabio and {Negro}, Michela and {Ng}, C.-Y. and {O'Dell}, Stephen L. and {Omodei}, Nicola and {Oppedisano}, Chiara and {Papitto}, Alessandro and {Peirson}, Abel Lawrence and {Perri}, Matteo and {Pesce-Rollins}, Melissa and {Petrucci}, Pierre-Olivier and {Pilia}, Maura and {Possenti}, Andrea and {Puccetti}, Simonetta and {Ramsey}, Brian D. and {Rankin}, John and {Ratheesh}, Ajay and {Roberts}, Oliver J. and {Romani}, Roger W. and {Sgr{\`o}}, Carmelo and {Slane}, Patrick and {Soffitta}, Paolo and {Spandre}, Gloria and {Swartz}, Douglas A. and {Tamagawa}, Toru and {Tavecchio}, Fabrizio and {Taverna}, Roberto and {Tawara}, Yuzuru and {Tennant}, Allyn F. and {Thomas}, Nicholas E. and {Tombesi}, Francesco and {Trois}, Alessio and {Turolla}, Roberto and {Vink}, Jacco and {Weisskopf}, Martin C. and {Wu}, Kinwah and {Xie}, Fei and {Zane}, Silvia},
        title = "{Complex rotational dynamics of the neutron star in Hercules X-1 revealed by X-ray polarization}",
      journal = {Nature Astronomy},
         year = 2024,
        month = aug,
       volume = {8},
        pages = {1047-1053},
          doi = {10.1038/s41550-024-02295-8},
       adsurl = {https://ui.adsabs.harvard.edu/abs/2024NatAs...8.1047H}
}

@ARTICLE{Mushtukov+22,
       author = {{Mushtukov}, Alexander and {Tsygankov}, Sergey},
        title = "{Accreting strongly magnetised neutron stars: X-ray Pulsars}",
      journal = {arXiv e-prints},
         year = 2022,
        month = apr,
          eid = {arXiv:2204.14185},
        pages = {arXiv:2204.14185},
          doi = {10.48550/arXiv.2204.14185},
archivePrefix = {arXiv},
       eprint = {2204.14185},
 primaryClass = {astro-ph.HE},
       adsurl = {https://ui.adsabs.harvard.edu/abs/2022arXiv220414185M}
}

@ARTICLE{Ghosh+79,
       author = {{Ghosh}, P. and {Lamb}, F.~K.},
        title = "{Accretion by rotating magnetic neutron stars. II. Radial and vertical structure of the transition zone in disk accretion.}",
      journal = {\apj},
         year = 1979,
        month = aug,
       volume = {232},
        pages = {259-276},
          doi = {10.1086/157285},
       adsurl = {https://ui.adsabs.harvard.edu/abs/1979ApJ...232..259G}
}

@ARTICLE{Lamb+73,
       author = {{Lamb}, F.~K. and {Pethick}, C.~J. and {Pines}, D.},
        title = "{A Model for Compact X-Ray Sources: Accretion by Rotating Magnetic Stars}",
      journal = {\apj},
         year = 1973,
        month = aug,
       volume = {184},
        pages = {271-290},
          doi = {10.1086/152325},
       adsurl = {https://ui.adsabs.harvard.edu/abs/1973ApJ...184..271L}
}

@ARTICLE{Basko+76,
       author = {{Basko}, M.~M. and {Sunyaev}, R.~A.},
        title = "{The limiting luminosity of accreting neutron stars with magnetic fields.}",
      journal = {\mnras},
         year = 1976,
        month = may,
       volume = {175},
        pages = {395-417},
          doi = {10.1093/mnras/175.2.395},
       adsurl = {https://ui.adsabs.harvard.edu/abs/1976MNRAS.175..395B}
}

@ARTICLE{Becker+07,
       author = {{Becker}, Peter A. and {Wolff}, Michael T.},
        title = "{Thermal and Bulk Comptonization in Accretion-powered X-Ray Pulsars}",
      journal = {\apj},
         year = 2007,
        month = jan,
       volume = {654},
       number = {1},
        pages = {435-457},
          doi = {10.1086/509108},
archivePrefix = {arXiv},
       eprint = {astro-ph/0609035},
 primaryClass = {astro-ph},
       adsurl = {https://ui.adsabs.harvard.edu/abs/2007ApJ...654..435B}
}

@ARTICLE{Ghosh+78,
       author = {{Ghosh}, P. and {Lamb}, F.~K.},
        title = "{Disk accretion by magnetic neutron stars.}",
      journal = {\apjl},
         year = 1978,
        month = jul,
       volume = {223},
        pages = {L83-L87},
          doi = {10.1086/182734},
       adsurl = {https://ui.adsabs.harvard.edu/abs/1978ApJ...223L..83G}
}

@ARTICLE{Brice+21,
       author = {{Brice}, Nabil and {Zane}, Silvia and {Turolla}, Roberto and {Wu}, Kinwah},
        title = "{Super-eddington emission from accreting, highly magnetized neutron stars with a multipolar magnetic field}",
      journal = {\mnras},
         year = 2021,
        month = jun,
       volume = {504},
       number = {1},
        pages = {701-715},
          doi = {10.1093/mnras/stab915},
archivePrefix = {arXiv},
       eprint = {2104.06138},
 primaryClass = {astro-ph.HE},
       adsurl = {https://ui.adsabs.harvard.edu/abs/2021MNRAS.504..701B}
}

@ARTICLE{Giacconi+71,
       author = {{Giacconi}, R. and {Gursky}, H. and {Kellogg}, E. and {Schreier}, E. and {Tananbaum}, H.},
        title = "{Discovery of Periodic X-Ray Pulsations in Centaurus X-3 from UHURU}",
      journal = {\apjl},
         year = 1971,
        month = jul,
       volume = {167},
        pages = {L67},
          doi = {10.1086/180762},
       adsurl = {https://ui.adsabs.harvard.edu/abs/1971ApJ...167L..67G}
}

@ARTICLE{Staubert+09,
       author = {{Staubert}, R. and {Klochkov}, D. and {Wilms}, J.},
        title = "{Updating the orbital ephemeris of Hercules X-1; rate of decay and eccentricity of the orbit}",
      journal = {\aap},
         year = 2009,
        month = jun,
       volume = {500},
       number = {2},
        pages = {883-889},
          doi = {10.1051/0004-6361/200911690},
archivePrefix = {arXiv},
       eprint = {0904.2307},
 primaryClass = {astro-ph.HE},
       adsurl = {https://ui.adsabs.harvard.edu/abs/2009A&A...500..883S}
}

@ARTICLE{Finger+09,
       author = {{Finger}, Mark H. and {Beklen}, Elif and {Narayana Bhat}, P. and {Paciesas}, William S. and {Connaughton}, Valerie and {Buckley}, David A.~H. and {Camero-Arranz}, Ascension and {Coe}, Malcolm J. and {Jenke}, Peter and {Kanbach}, Gottfried and {Negueruela}, Ignacio and {Wilson-Hodge}, Colleen A.},
        title = "{Long-term Monitoring of Accreting Pulsars with Fermi GBM}",
      journal = {arXiv e-prints},
         year = 2009,
        month = dec,
          eid = {arXiv:0912.3847},
        pages = {arXiv:0912.3847},
          doi = {10.48550/arXiv.0912.3847},
archivePrefix = {arXiv},
       eprint = {0912.3847},
 primaryClass = {astro-ph.HE},
       adsurl = {https://ui.adsabs.harvard.edu/abs/2009arXiv0912.3847F}
}

@ARTICLE{Stingray+19,
          author = {{Huppenkothen}, Daniela and {Bachetti}, Matteo and
                    {Stevens}, Abigail L. and {Migliari}, Simone and {Balm}, Paul and
                    {Hammad}, Omar and {Khan}, Usman Mahmood and {Mishra}, Himanshu and
                    {Rashid}, Haroon and {Sharma}, Swapnil and {Martinez Ribeiro}, Evandro and
                    {Valles Blanco}, Ricardo},
          title = "{Stingray: A Modern Python Library for Spectral Timing}",
          journal = {apj},
          year = 2019,
          month = aug,
          volume = {881},
          number = {1},
          eid = {39},
          pages = {39},
          doi = {10.3847/1538-4357/ab258d},
          archivePrefix = {arXiv},
          eprint = {1901.07681},
          primaryClass = {astro-ph.IM},
          adsurl = {https://ui.adsabs.harvard.edu/abs/2019ApJ...881...39H}
        }

@ARTICLE{Huppenkothen+19,
          doi = {10.21105/joss.01393},
          url = {https://doi.org/10.21105/joss.01393},
          year = {2019},
          publisher = {The Open Journal},
          volume = {4},
          number = {38},
          pages = {1393},
          author = {Daniela Huppenkothen and Matteo Bachetti and Abigail Stevens and Simone Migliari and Paul Balm and Omar Hammad and Usman Mahmood Khan and Himanshu Mishra and Haroon Rashid and Swapnil Sharma and Evandro Martinez Ribeiro and Ricardo Valles Blanco},
          title = {stingray: A modern Python library for spectral timing},
          journal = {Journal of Open Source Software}
        }

@ARTICLE{Ferrigno+09,
       author = {{Ferrigno}, C. and {Becker}, P.~A. and {Segreto}, A. and {Mineo}, T. and {Santangelo}, A.},
        title = "{Study of the accreting pulsar 4U 0115+63 using a bulk and thermal Comptonization model}",
      journal = {\aap},
         year = 2009,
        month = may,
       volume = {498},
       number = {3},
        pages = {825-836},
          doi = {10.1051/0004-6361/200809373},
archivePrefix = {arXiv},
       eprint = {0902.4392},
 primaryClass = {astro-ph.HE},
       adsurl = {https://ui.adsabs.harvard.edu/abs/2009A&A...498..825F}
}

@ARTICLE{Kallman+01,
       author = {{Kallman}, T. and {Bautista}, M.},
        title = "{Photoionization and High-Density Gas}",
      journal = {\apjs},
         year = 2001,
        month = mar,
       volume = {133},
       number = {1},
        pages = {221-253},
          doi = {10.1086/319184},
       adsurl = {https://ui.adsabs.harvard.edu/abs/2001ApJS..133..221K}
}

@ARTICLE{Parker+09,
       author = {{Parker}, M.~L. and {Longinotti}, A.~L. and {Schartel}, N. and {Grupe}, D. and {Komossa}, S. and {Kriss}, G. and {Fabian}, A.~C. and {Gallo}, L. and {Harrison}, F.~A. and {Jiang}, J. and {Kara}, E. and {Krongold}, Y. and {Matzeu}, G.~A. and {Pinto}, C. and {Santos-Lle{\'o}}, M.},
        title = "{The nuclear environment of the NLS1 Mrk 335: Obscuration of the X-ray line emission by a variable outflow}",
      journal = {\mnras},
         year = 2019,
        month = nov,
       volume = {490},
       number = {1},
        pages = {683-697},
          doi = {10.1093/mnras/stz2566},
archivePrefix = {arXiv},
       eprint = {1909.04924},
 primaryClass = {astro-ph.HE},
       adsurl = {https://ui.adsabs.harvard.edu/abs/2019MNRAS.490..683P}
}

@ARTICLE{Wilms+00,
       author = {{Wilms}, J. and {Allen}, A. and {McCray}, R.},
        title = "{On the Absorption of X-Rays in the Interstellar Medium}",
      journal = {\apj},
         year = 2000,
        month = oct,
       volume = {542},
       number = {2},
        pages = {914-924},
          doi = {10.1086/317016},
archivePrefix = {arXiv},
       eprint = {astro-ph/0008425},
 primaryClass = {astro-ph},
       adsurl = {https://ui.adsabs.harvard.edu/abs/2000ApJ...542..914W}
}

@ARTICLE{Thalhammer+21,
       author = {{Thalhammer}, Philipp and {Bissinger}, Matthias and {Ballhausen}, Ralf and {Pottschmidt}, Katja and {Wolff}, Michael T. and {Stierhof}, Jakob and {Sokolova-Lapa}, Ekaterina and {F{\"u}rst}, Felix and {Malacaria}, Christian and {Gottlieb}, Amy and {Marcu-Cheatham}, Diana M. and {Becker}, Peter A. and {Wilms}, J{\"o}rn},
        title = "{Fitting strategies of accretion column models and application to the broadband spectrum of Cen X-3}",
      journal = {\aap},
         year = 2021,
        month = dec,
       volume = {656},
          eid = {A105},
        pages = {A105},
          doi = {10.1051/0004-6361/202140582},
archivePrefix = {arXiv},
       eprint = {2109.14565},
 primaryClass = {astro-ph.HE},
       adsurl = {https://ui.adsabs.harvard.edu/abs/2021A&A...656A.105T}
}

@ARTICLE{Wolff+16,
       author = {{Wolff}, Michael T. and {Becker}, Peter A. and {Gottlieb}, Amy M. and {F{\"u}rst}, Felix and {Hemphill}, Paul B. and {Marcu-Cheatham}, Diana M. and {Pottschmidt}, Katja and {Schwarm}, Fritz-Walter and {Wilms}, J{\"o}rn and {Wood}, Kent S.},
        title = "{The NuSTAR X-Ray Spectrum of Hercules X-1: A Radiation-dominated Radiative Shock}",
      journal = {\apj},
         year = 2016,
        month = nov,
       volume = {831},
       number = {2},
          eid = {194},
        pages = {194},
          doi = {10.3847/0004-637X/831/2/194},
archivePrefix = {arXiv},
       eprint = {1608.08978},
 primaryClass = {astro-ph.HE},
       adsurl = {https://ui.adsabs.harvard.edu/abs/2016ApJ...831..194W}
}

@ARTICLE{Fuerst+13,
       author = {{F{\"u}rst}, Felix and {Grefenstette}, Brian W. and {Staubert}, R{\"u}diger and {Tomsick}, John A. and {Bachetti}, Matteo and {Barret}, Didier and {Bellm}, Eric C. and {Boggs}, Steven E. and {Chenevez}, Jerome and {Christensen}, Finn E. and {Craig}, William W. and {Hailey}, Charles J. and {Harrison}, Fiona and {Klochkov}, Dmitry and {Madsen}, Kristin K. and {Pottschmidt}, Katja and {Stern}, Daniel and {Walton}, Dominic J. and {Wilms}, J{\"o}rn and {Zhang}, William},
        title = "{The Smooth Cyclotron Line in Her X-1 as Seen with Nuclear Spectroscopic Telescope Array}",
      journal = {\apj},
         year = 2013,
        month = dec,
       volume = {779},
       number = {1},
          eid = {69},
        pages = {69},
          doi = {10.1088/0004-637X/779/1/69},
       adsurl = {https://ui.adsabs.harvard.edu/abs/2013ApJ...779...69F}
}

@ARTICLE{Asami+14,
       author = {{Asami}, Fumi and {Enoto}, Teruaki and {Iwakiri}, Wataru and {Yamada}, Shin'ya and {Tamagawa}, Toru and {Mihara}, Tatehiro and {Nagase}, Fumiaki},
        title = "{Broad-band spectroscopy of Hercules X-1 with Suzaku}",
      journal = {\pasj},
         year = 2014,
        month = apr,
       volume = {66},
       number = {2},
          eid = {44},
        pages = {44},
          doi = {10.1093/pasj/psu005},
       adsurl = {https://ui.adsabs.harvard.edu/abs/2014PASJ...66...44A}
}

@BOOK{Ruder+94,
       author = {{Ruder}, Hanns and {Wunner}, G{\"u}nter and {Herold}, Heinz and {Geyer}, Florian},
        title = "{Atoms in Strong Magnetic Fields. Quantum Mechanical Treatment and Applications in Astrophysics and Quantum Chaos}",
         year = 1994,
       adsurl = {https://ui.adsabs.harvard.edu/abs/1994asmf.book.....R}
}

@ARTICLE{Fabian+89,
       author = {{Fabian}, A.~C. and {Rees}, M.~J. and {Stella}, L. and {White}, N.~E.},
        title = "{X-ray fluorescence from the inner disc in Cygnus X-1.}",
      journal = {\mnras},
         year = 1989,
        month = may,
       volume = {238},
        pages = {729-736},
          doi = {10.1093/mnras/238.3.729},
       adsurl = {https://ui.adsabs.harvard.edu/abs/1989MNRAS.238..729F}
}

@ARTICLE{Garcia+14,
       author = {{Garc{\'\i}a}, J. and {Dauser}, T. and {Lohfink}, A. and {Kallman}, T.~R. and {Steiner}, J.~F. and {McClintock}, J.~E. and {Brenneman}, L. and {Wilms}, J. and {Eikmann}, W. and {Reynolds}, C.~S. and {Tombesi}, F.},
        title = "{Improved Reflection Models of Black Hole Accretion Disks: Treating the Angular Distribution of X-Rays}",
      journal = {\apj},
         year = 2014,
        month = feb,
       volume = {782},
       number = {2},
          eid = {76},
        pages = {76},
          doi = {10.1088/0004-637X/782/2/76},
archivePrefix = {arXiv},
       eprint = {1312.3231},
 primaryClass = {astro-ph.HE},
       adsurl = {https://ui.adsabs.harvard.edu/abs/2014ApJ...782...76G}
}

@ARTICLE{Ludlam+24,
       author = {{Ludlam}, Renee M.},
        title = "{Reflecting on accretion in neutron star low-mass X-ray binaries}",
      journal = {\apss},
         year = 2024,
        month = jan,
       volume = {369},
       number = {1},
          eid = {16},
        pages = {16},
          doi = {10.1007/s10509-024-04281-y},
archivePrefix = {arXiv},
       eprint = {2401.15787},
 primaryClass = {astro-ph.HE},
       adsurl = {https://ui.adsabs.harvard.edu/abs/2024Ap&SS.369...16L}
}

@ARTICLE{Ludlam+25,
       author = {{Ludlam}, R.~M. and {Miller}, J.~M. and {Cackett}, E.~M. and {Garc{\'\i}a}, J.~A.},
        title = "{XRISM Resolves Relativistic Effects from the Innermost Accretion Disk in Serpens X-1}",
      journal = {\apjl},
         year = 2025,
        month = nov,
       volume = {993},
       number = {2},
          eid = {L41},
        pages = {L41},
          doi = {10.3847/2041-8213/ae13a5},
archivePrefix = {arXiv},
       eprint = {2510.13739},
 primaryClass = {astro-ph.HE},
       adsurl = {https://ui.adsabs.harvard.edu/abs/2025ApJ...993L..41L}
}

@ARTICLE{Wilson-Hodge+18,
       author = {{Wilson-Hodge}, Colleen A. and {Malacaria}, Christian and {Jenke}, Peter A. and {Jaisawal}, Gaurava K. and {Kerr}, Matthew and {Wolff}, Michael T. and {Arzoumanian}, Zaven and {Chakrabarty}, Deepto and {Doty}, John P. and {Gendreau}, Keith C. and {Guillot}, Sebastien and {Ho}, Wynn C.~G. and {LaMarr}, Beverly and {Markwardt}, Craig B. and {{\"O}zel}, Feryal and {Prigozhin}, Gregory Y. and {Ray}, Paul S. and {Ramos-Lerate}, Mercedes and {Remillard}, Ronald A. and {Strohmayer}, Tod E. and {Vezie}, Michael L. and {Wood}, Kent S. and {NICER Science Team}},
        title = "{NICER and Fermi GBM Observations of the First Galactic Ultraluminous X-Ray Pulsar Swift J0243.6+6124}",
      journal = {\apj},
         year = 2018,
        month = aug,
       volume = {863},
       number = {1},
          eid = {9},
        pages = {9},
          doi = {10.3847/1538-4357/aace60},
archivePrefix = {arXiv},
       eprint = {1806.10094},
 primaryClass = {astro-ph.HE},
       adsurl = {https://ui.adsabs.harvard.edu/abs/2018ApJ...863....9W}
}

@ARTICLE{Jaisawal+19,
       author = {{Jaisawal}, Gaurava K. and {Wilson-Hodge}, Colleen A. and {Fabian}, Andrew C. and {Naik}, Sachindra and {Chakrabarty}, Deepto and {Kretschmar}, Peter and {Ballantyne}, David R. and {Ludlam}, Renee M. and {Chenevez}, J{\'e}r{\^o}me and {Altamirano}, Diego and {Arzoumanian}, Zaven and {F{\"u}rst}, Felix and {Gendreau}, Keith C. and {Guillot}, Sebastien and {Malacaria}, Christian and {Miller}, Jon M. and {Stevens}, Abigail L. and {Wolff}, Michael T.},
        title = "{An Evolving Broad Iron Line from the First Galactic Ultraluminous X-Ray Pulsar Swift J0243.6+6124}",
      journal = {\apj},
         year = 2019,
        month = nov,
       volume = {885},
       number = {1},
          eid = {18},
        pages = {18},
          doi = {10.3847/1538-4357/ab4595},
archivePrefix = {arXiv},
       eprint = {1909.07338},
 primaryClass = {astro-ph.HE},
       adsurl = {https://ui.adsabs.harvard.edu/abs/2019ApJ...885...18J}
}

@ARTICLE{Blum+00,
       author = {{Blum}, S. and {Kraus}, U.},
        title = "{Analyzing X-Ray Pulsar Profiles: Geometry and Beam Pattern of Hercules X-1}",
      journal = {\apj},
         year = 2000,
        month = feb,
       volume = {529},
       number = {2},
        pages = {968-977},
          doi = {10.1086/308308},
archivePrefix = {arXiv},
       eprint = {astro-ph/9909449},
 primaryClass = {astro-ph},
       adsurl = {https://ui.adsabs.harvard.edu/abs/2000ApJ...529..968B}
}

@ARTICLE{Chakraborty+24,
       author = {{Chakraborty}, Priyanka and {Ferland}, Gary and {Fabian}, Andrew and {Sarkar}, Arnab and {Ludlam}, Renee and {Bianchi}, Stefano and {Hall}, Hayden and {Kosec}, Peter},
        title = "{Physics of 1 keV line in X-ray binaries}",
      journal = {arXiv e-prints},
         year = 2024,
        month = jul,
          eid = {arXiv:2407.02360},
        pages = {arXiv:2407.02360},
          doi = {10.48550/arXiv.2407.02360},
archivePrefix = {arXiv},
       eprint = {2407.02360},
 primaryClass = {astro-ph.HE},
       adsurl = {https://ui.adsabs.harvard.edu/abs/2024arXiv240702360C}
}

@ARTICLE{Staubert+09a,
       author = {{Staubert}, R. and {Klochkov}, D. and {Postnov}, K. and {Shakura}, N. and {Wilms}, J. and {Rothschild}, R.~E.},
        title = "{Two \raisebox{-0.5ex}\textasciitilde35 day clocks in Hercules X-1: evidence for neutron star free precession}",
      journal = {\aap},
         year = 2009,
        month = feb,
       volume = {494},
       number = {3},
        pages = {1025-1030},
          doi = {10.1051/0004-6361:200810743},
archivePrefix = {arXiv},
       eprint = {0811.4045},
 primaryClass = {astro-ph},
       adsurl = {https://ui.adsabs.harvard.edu/abs/2009A&A...494.1025S}
}

@ARTICLE{Garcia+13,
       author = {{Garc{\'\i}a}, J. and {Dauser}, T. and {Reynolds}, C.~S. and {Kallman}, T.~R. and {McClintock}, J.~E. and {Wilms}, J. and {Eikmann}, W.},
        title = "{X-Ray Reflected Spectra from Accretion Disk Models. III. A Complete Grid of Ionized Reflection Calculations}",
      journal = {\apj},
         year = 2013,
        month = may,
       volume = {768},
       number = {2},
          eid = {146},
        pages = {146},
          doi = {10.1088/0004-637X/768/2/146},
archivePrefix = {arXiv},
       eprint = {1303.2112},
 primaryClass = {astro-ph.HE},
       adsurl = {https://ui.adsabs.harvard.edu/abs/2013ApJ...768..146G}
}

@ARTICLE{Ji+09,
       author = {{Ji}, L. and {Schulz}, N. and {Nowak}, M. and {Marshall}, H.~L. and {Kallman}, T.},
        title = "{The Photoionized Accretion Disk in Her X-1}",
      journal = {\apj},
         year = 2009,
        month = aug,
       volume = {700},
       number = {2},
        pages = {977-988},
          doi = {10.1088/0004-637X/700/2/977},
archivePrefix = {arXiv},
       eprint = {0905.3773},
 primaryClass = {astro-ph.HE},
       adsurl = {https://ui.adsabs.harvard.edu/abs/2009ApJ...700..977J}
}

@ARTICLE{Leahy+15,
       author = {{Leahy}, D.~A.},
        title = "{Hercules X-1: Using Eclipse to Measure the X-Ray Corona}",
      journal = {\apj},
         year = 2015,
        month = feb,
       volume = {800},
       number = {1},
          eid = {32},
        pages = {32},
          doi = {10.1088/0004-637X/800/1/32},
       adsurl = {https://ui.adsabs.harvard.edu/abs/2015ApJ...800...32L}
}

@ARTICLE{Garcia+22,
       author = {{Garc{\'\i}a}, Javier A. and {Dauser}, Thomas and {Ludlam}, Renee and {Parker}, Michael and {Fabian}, Andrew and {Harrison}, Fiona A. and {Wilms}, J{\"o}rn},
        title = "{Relativistic X-Ray Reflection Models for Accreting Neutron Stars}",
      journal = {\apj},
         year = 2022,
        month = feb,
       volume = {926},
       number = {1},
          eid = {13},
        pages = {13},
          doi = {10.3847/1538-4357/ac3cb7},
archivePrefix = {arXiv},
       eprint = {2111.12838},
 primaryClass = {astro-ph.HE},
       adsurl = {https://ui.adsabs.harvard.edu/abs/2022ApJ...926...13G}
}

@ARTICLE{Mushtukov+24,
       author = {{Mushtukov}, Alexander A. and {Weng}, Albert and {Tsygankov}, Sergey S. and {Mereminskiy}, Ilya A.},
        title = "{Flickering pulsations in bright X-ray pulsars: the evidence of gravitationally lensed and eclipsed accretion column}",
      journal = {\mnras},
         year = 2024,
        month = may,
       volume = {530},
       number = {3},
        pages = {3051-3058},
          doi = {10.1093/mnras/stae967},
archivePrefix = {arXiv},
       eprint = {2404.04137},
 primaryClass = {astro-ph.HE},
       adsurl = {https://ui.adsabs.harvard.edu/abs/2024MNRAS.530.3051M}
}

@ARTICLE{Groger+26,
       author = {{Groger}, John and {Paerels}, Frits and {Bogdanov}, Slavko and {Gotthelf}, Eric V. and {Helfand}, David J. and {Hubeny}, Ivan and {Lanz}, Thierry and {Gomez}, Thomas A.},
        title = "{Evidence for Atomic Absorption Features in the High Resolution X-ray Spectrum of the Neutron Star in Puppis A}",
      journal = {arXiv e-prints},
         year = 2026,
        month = mar,
          eid = {arXiv:2603.14146},
        pages = {arXiv:2603.14146},
          doi = {10.48550/arXiv.2603.14146},
archivePrefix = {arXiv},
       eprint = {2603.14146},
 primaryClass = {astro-ph.HE},
       adsurl = {https://ui.adsabs.harvard.edu/abs/2026arXiv260314146G}
}

@ARTICLE{Schandl+94,
       author = {{Schandl}, S. and {Meyer}, F.},
        title = "{Herculis X-1: coronal winds producing the tilted shape of the accretion disk.}",
      journal = {\aap},
         year = 1994,
        month = sep,
       volume = {289},
        pages = {149-161},
       adsurl = {https://ui.adsabs.harvard.edu/abs/1994A&A...289..149S}
}

@ARTICLE{Ogilvie+01,
       author = {{Ogilvie}, G.~I. and {Dubus}, G.},
        title = "{Precessing warped accretion discs in X-ray binaries}",
      journal = {\mnras},
         year = 2001,
        month = feb,
       volume = {320},
       number = {4},
        pages = {485-503},
          doi = {10.1046/j.1365-8711.2001.04011.x},
archivePrefix = {arXiv},
       eprint = {astro-ph/0009264},
 primaryClass = {astro-ph},
       adsurl = {https://ui.adsabs.harvard.edu/abs/2001MNRAS.320..485O}
}

@ARTICLE{Link+07,
       author = {{Link}, Bennett},
        title = "{Precession as a probe of the neutron star interior}",
      journal = {\apss},
         year = 2007,
        month = apr,
       volume = {308},
       number = {1-4},
        pages = {435-441},
          doi = {10.1007/s10509-007-9315-0},
       adsurl = {https://ui.adsabs.harvard.edu/abs/2007Ap&SS.308..435L}
}

@ARTICLE{Kong+22,
       author = {{Kong}, Ling-Da and {Zhang}, Shu and {Zhang}, Shuang-Nan and {Ji}, Long and {Doroshenko}, Victor and {Santangelo}, Andrea and {Chen}, Yu-Peng and {Lu}, Fang-Jun and {Ge}, Ming-Yu and {Wang}, Peng-Ju and {Tao}, Lian and {Qu}, Jin-Lu and {Li}, Ti-Pei and {Liu}, Cong-Zhan and {Liao}, Jin-Yuan and {Chang}, Zhi and {Peng}, Jing-Qiang and {Shui}, Qing-Cang},
        title = "{Insight-HXMT Discovery of the Highest-energy CRSF from the First Galactic Ultraluminous X-Ray Pulsar Swift J0243.6+6124}",
      journal = {\apjl},
         year = 2022,
        month = jul,
       volume = {933},
       number = {1},
          eid = {L3},
        pages = {L3},
          doi = {10.3847/2041-8213/ac7711},
archivePrefix = {arXiv},
       eprint = {2206.04283},
 primaryClass = {astro-ph.HE},
       adsurl = {https://ui.adsabs.harvard.edu/abs/2022ApJ...933L...3K}
}

@ARTICLE{Mandal+23,
       author = {{Mandal}, Manoj and {Sharma}, Rahul and {Pal}, Sabyasachi and {Jaisawal}, G.~K. and {Gendreau}, Keith C. and {Ng}, Mason and {Sanna}, Andrea and {Malacaria}, Christian and {Tombesi}, Francesco and {Ferrara}, E.~C. and {Markwardt}, Craig B. and {Wolff}, Michael T. and {Coley}, Joel B.},
        title = "{Probing spectral and timing properties of the X-ray pulsar RX J0440.9 + 4431 in the giant outburst of 2022-2023}",
      journal = {\mnras},
         year = 2023,
        month = nov,
       volume = {526},
       number = {1},
        pages = {771-781},
          doi = {10.1093/mnras/stad2767},
archivePrefix = {arXiv},
       eprint = {2306.08083},
 primaryClass = {astro-ph.HE},
       adsurl = {https://ui.adsabs.harvard.edu/abs/2023MNRAS.526..771M}
}

@ARTICLE{Tarter+69,
       author = {{Tarter}, C. Bruce and {Tucker}, Wallace H. and {Salpeter}, Edwin E.},
        title = "{The Interaction of X-Ray Sources with Optically Thin Environments}",
      journal = {\apj},
         year = 1969,
        month = jun,
       volume = {156},
        pages = {943},
          doi = {10.1086/150026},
       adsurl = {https://ui.adsabs.harvard.edu/abs/1969ApJ...156..943T}
}

@ARTICLE{Ross+78,
       author = {{Ross}, R.~R. and {Weaver}, R. and {McCray}, R.},
        title = "{The Comptonization of iron X-ray features in compact X-ray sources.}",
      journal = {\apj},
         year = 1978,
        month = jan,
       volume = {219},
        pages = {292-299},
          doi = {10.1086/155776},
       adsurl = {https://ui.adsabs.harvard.edu/abs/1978ApJ...219..292R}
}

@ARTICLE{Eckart+18,
       author = {{Eckart}, Megan E. and {Adams}, Joseph S. and {Boyce}, Kevin R. and {Brown}, Gregory V. and {Chiao}, Meng P. and {Fujimoto}, Ryuichi and {Haas}, Daniel and {den Herder}, Jan-Willem and {Hoshino}, Akio and {Ishisaki}, Yoshitaka and {Kilbourne}, Caroline A. and {Kitamoto}, Shunji and {Leutenegger}, Maurice A. and {McCammon}, Dan and {Mitsuda}, Kazuhisa and {Porter}, F. Scott and {Sato}, Kosuke and {Sawada}, Makoto and {Seta}, Hiromi and {Sneiderman}, Gary A. and {Szymkowiak}, Andrew E. and {Takei}, Yoh and {Tashiro}, Makoto S. and {Tsujimoto}, Masahiro and {de Vries}, Cor P. and {Watanabe}, Tomomi and {Yamada}, Shinya and {Yamasaki}, Noriko Y.},
        title = "{Ground calibration of the Astro-H (Hitomi) soft x-ray spectrometer}",
      journal = {Journal of Astronomical Telescopes, Instruments, and Systems},
         year = 2018,
        month = apr,
       volume = {4},
          eid = {021406},
        pages = {021406},
          doi = {10.1117/1.JATIS.4.2.021406},
       adsurl = {https://ui.adsabs.harvard.edu/abs/2018JATIS...4b1406E}
}

@ARTICLE{Becker+98,
       author = {{Becker}, Peter A.},
        title = "{Dynamical Structure of Radiation-dominated Pulsar Accretion Shocks}",
      journal = {\apj},
         year = 1998,
        month = may,
       volume = {498},
       number = {2},
        pages = {790-801},
          doi = {10.1086/305568},
       adsurl = {https://ui.adsabs.harvard.edu/abs/1998ApJ...498..790B}
}

@ARTICLE{Tanaka+95,
       author = {{Tanaka}, Y. and {Nandra}, K. and {Fabian}, A.~C. and {Inoue}, H. and {Otani}, C. and {Dotani}, T. and {Hayashida}, K. and {Iwasawa}, K. and {Kii}, T. and {Kunieda}, H. and {Makino}, F. and {Matsuoka}, M.},
        title = "{Gravitationally redshifted emission implying an accretion disk and massive black hole in the active galaxy MCG-6-30-15}",
      journal = {\nat},
         year = 1995,
        month = jun,
       volume = {375},
       number = {6533},
        pages = {659-661},
          doi = {10.1038/375659a0},
       adsurl = {https://ui.adsabs.harvard.edu/abs/1995Natur.375..659T}
}

@ARTICLE{Becker+22,
       author = {{Becker}, Peter A. and {Wolff}, Michael T.},
        title = "{A Generalized Analytical Model for Thermal and Bulk Comptonization in Accretion-powered X-Ray Pulsars}",
      journal = {\apj},
         year = 2022,
        month = nov,
       volume = {939},
       number = {2},
          eid = {67},
        pages = {67},
          doi = {10.3847/1538-4357/ac8d95},
archivePrefix = {arXiv},
       eprint = {2211.13894},
 primaryClass = {astro-ph.HE},
       adsurl = {https://ui.adsabs.harvard.edu/abs/2022ApJ...939...67B}
}
\bibliographystyle{aasjournalv7}



\end{document}